\documentclass{article}

\usepackage{arxiv}

\usepackage[utf8]{inputenc} 
\usepackage[T1]{fontenc}    
\usepackage{hyperref}       
\usepackage{url}            
\usepackage{booktabs}       
\usepackage{amsfonts}       
\usepackage{nicefrac}       
\usepackage{microtype}      
\usepackage{lipsum}
\usepackage{fancyhdr}       
\usepackage{graphicx}       
\graphicspath{{media/}} 
\usepackage{tikz}
\usepackage{pgfplots}
\usetikzlibrary{patterns}
\usepackage{balance}
\usepackage{subcaption}
\usepackage{tabularx}
\usepackage{amssymb}
\usepackage{graphicx}
\usepackage{caption}
\usetikzlibrary{shapes.geometric}
\usetikzlibrary{arrows.meta}
\newcommand{\cmark}{\ding{51}}%
\newcommand{\xmark}{\ding{55}}
\usepackage{array} 
\usepackage{amssymb}
\usepackage{pifont}

\title{Deep Learning Fusion For Effective Malware Detection: Leveraging Visual Features
}

\author{%
  Jahez Abraham Johny\textsuperscript{1},
  Vinod P.\textsuperscript{2,4},
  Asmitha K. A. \textsuperscript{2},
  G. Radhamani \textsuperscript{1},
  Rafidha Rehiman K. A.\textsuperscript{2}
  Mauro Conti\textsuperscript{4}
}
\affiliation{%
    \bf{\textsuperscript{1}} Department of Computer Science, Dr. G. R. Damodaran College of Science, India \\
   \bf{\textsuperscript{2}} Department of Computer Applications, Cochin University of Science and Technology, India \\
   \bf{\textsuperscript{4}} Department of Mathematics, University of Padua, Italy \\
  \bigskip
  Corresponding author:
  \texttt{22mca018@grd.edu.in};
  \texttt{vinod.puthuvath@unipd.it}; \texttt{vinod.p@cusat.ac.in}
  \texttt{asmitha@pg.cusat.ac.in};
  \texttt{radhamani@grd.edu.in};
  \texttt{rafidharhimanka@cusat.ac.in};
  \texttt{mauro.conti@unipd.it};
}

\begin{document}
\maketitle

\begin{abstract}
Malware has become a formidable threat as it has been growing exponentially in number and sophistication, thus, it is imperative to have a solution that is easy to implement, reliable, and effective. While recent research has introduced deep learning multi-feature fusion algorithms, they lack a proper explanation. In this work, we investigate the power of fusing Convolutional Neural Network models trained on different modalities of a malware executable. We are proposing a novel multimodal fusion algorithm, leveraging three different visual malware features: Grayscale Image, Entropy Graph, and SimHash Image, with which we conducted exhaustive experiments independently on each feature and combinations of all three of them using fusion operators such as average, maximum, add, and concatenate for effective malware detection and classification.
The proposed strategy has a detection rate of 1.00 (on a scale of 0-1) in identifying malware in the given dataset. We explained its interpretability with visualization techniques such as t-SNE and Grad-CAM. Experimental results show the model works even for a highly imbalanced dataset. We also assessed the effectiveness of the proposed method on obfuscated malware and achieved state-of-the-art results. The proposed methodology is more reliable as our findings prove VGG16 model can detect and classify malware in a matter of seconds in real-time.
\end{abstract}


\section{Introduction}
\label{sec:introduction}
Malware, a malicious software, aims to steal sensitive digital information. As the internet and its technologies evolve rapidly, the number of vulnerabilities exploited by malware has surged alarmingly. 
The threat reports from AV-Test \footnote{https://www.av-test.org/en/statistics/malware/} institute show that it detects 450,000 novel malware and potentially unwanted applications (PUA) daily. Shockingly, in January 2024 alone, a staggering 839,859,176 Windows malware instances were reported. Confronted with the rapid evolution of malware and the impracticality and inconsistency of manual analysis, antivirus companies and security organizations are turning to machine learning and artificial intelligence to accelerate the detection process. As a result, traditional machine learning techniques, such as Decision Trees, Support Vector Machines, Na\"ive Bayes, XGBoost, and Random Forest, have become prevalent in malware identification and classification. However, a significant drawback of these methods is the substantial computational resources needed for feature engineering and the intricate process of handling large datasets, thus raising questions on the usability of such solutions. 
\par Malware detection/classification systems typically employ two distinct approaches: static and dynamic analysis. Static analysis scrutinizes the disassembled pcode, executable header, malware assembly programs, strings, and other structural elements to extract relevant features. Dynamic analysis, in contrast, observes the behavior of malware by analyzing its system calls, network communication, memory usage, and invoked processes. Both static and dynamic feature extraction can be intricate processes. Static analysis requires security professionals with a thorough understanding of malicious file structures. However, static analysis is ineffective in detecting crucial malicious code characteristics if the malware employs compression or encryption techniques. In contrast, dynamic analysis provides accurate insights into malware behavior only if the execution path that triggers malicious actions is systematically analyzed, making it vulnerable to evasive strategies employed by malware programs. Additionally, dynamic analysis is a relatively time-consuming process.
\par Reusing the core code generates new variants of a given malware family, which defines a collection of malicious binaries sharing similar features inherited from the base malware. Researchers noticed structural similarity in a specific class of malware executables and visualized the binary files into images, thereby creating an alternative method for classifying malicious code known as malware visualization. One of the initial and proposed works for classifying malware into its respective family was postulated by authors in~\cite{nataraj2011malware}. These researchers transformed executable files into grayscale images to extract visual features for malware categorization. With the quick advancement of Deep Learning architectures, particularly Convolutional Neural Networks (CNNs), the field of malware detection using visual features has gained significant traction. CNNs are particularly adept at identifying patterns and characteristics in visual data, making them an effective tool for classifying malware based on its texture and layout. 
\par Considerable research has been carried out in the area of malware detection using visual features, encompassing three main approaches: (i) \textit{Malware Binary Transformation:} techniques to convert malware binaries into different forms of visual representations, including grayscale images, RGB images, and Markov images~\cite{yuan2020byte}. These transformations aim to capture the inherent structural similarities and patterns within malware code, (ii) \textit{Feature Extraction:} use of popular feature extraction techniques include Local Binary Pattern~(LBP), Scale-Invariant Feature Transform~(SIFT), Graylevel co-occurrence matrix~(GLCM),  and Global Image Signature~(GIST) features\cite{humeau2019texture}, and (iii) \textit{Deep Learning Architecture Development:}  to particularly adept at extracting intricate patterns and features from complex visual data, making them well-suited for analyzing malware images, researchers are developing deep learning models that can accurately classify new malware samples with high detection rates. To further enhance the detection of new malware strains, researchers have explored fusing the outputs of multiple deep learning models with other machine learning and deep learning architectures\cite{martin2019android}\cite{gibert2020hydra}. This approach aims to combine the strengths of different models to achieve a more comprehensive and robust detection system. Additionally, the amalgamation of different modalities of features extracted from malware images is being investigated. 
\par Presently, researchers have not thoroughly investigated the combination of various deep learning models, particularly concerning the incorporation of images from diverse information sources of malware, which could notably enhance detection performance.
Further, the significance of the fusion/merging operator is not experimentally substantiated. Therefore, we believe that enhancing the malware classification system relying on visual traits requires addressing the following three criteria: (i) accurately assessing features unique to malware samples and those indicating similarity to other samples within the same family, (ii) ensuring the feature extraction time closely align with real-time antivirus detection, and (iii) simplifying the feature extraction phase with minimal involvement from security engineers. Our proposed study answers the following two research questions: 
\begin{itemize}
    \item \textbf{$RQ_1$:} What is the impact of image type variation on the performance of VGG16 models in malware classification, particularly focusing on grayscale, entropy graph, and simhash images?
    \item \textbf{$RQ_2$:} What is the impact of different fusion operators (e.g., average, maximum, add, concatenate) on the performance of the multimodal fusion algorithm for malware classification, particularly in terms of F1-score?
    \item \textbf{$RQ_3$:} What are the prominent features or patterns that facilitate the accurate classification of malware into families?
\end{itemize} 

To explore the aforementioned research query, we conducted a comprehensive examination involving three distinct types of images generated from a malware binary within the commonly utilized benchmark dataset, BIG2015. These images, specifically grayscale~(GH), entropygraph~(EG), and simhash~(SH) were thoroughly investigated to assess their impact on malware categorization. Additionally, employing various fusion methods such as \texttt{add}, \texttt{max}, \texttt{avg}, and \texttt{concatenate}(\textit{i.e.} the different ensembling techniques available in the Keras library) with VGG16 models, we conducted extensive experiments to determine the top-notch technique for integrating diverse modalities and enhancing malware classification. The contribution of the article can be summarized as below:

\begin{itemize}
    \item We devise a multimodal image exploration approach for malware categorization, utilizing grayscale, entropy graph, and simhash images to achieve enhanced detection accuracy and efficiency.
    
    \item We evaluate the effectiveness of different fusion methods, such as addition, maximum, average, and concatenation, to determine the optimal approach for merging multimodal images and enhancing malware classification accuracy.
    
    \item We design a malware classification system using a benchmark dataset with VGG16 model that improves categorization accuracy using multimodal image exploration techniques.

    \item We enhance the interpretability of decision-making of the proposed model utilizing Grad-CAM by highlighting crucial regions accounted for classification decisions and t-SNE for exploring clusters and trends. 
    \end{itemize}

\par The rest of the article is organized as follows. Section~\ref{sec:relatedwork} presents an overview of current malware classification methods. Section~\ref{sec:proposedmethod} details our proposed method. Section~\ref{sec:experimentsandresults} presents and compares experimental results obtained using our method to those of other studies. Finally, Section~\ref{sec:conclusion} summarizes the article's contributions. 
\section{Related work}
\label{sec:relatedwork}
Recently, visualization-based malware analysis has gained significant attention~\cite{vasan2020imcfn}~\cite{ni2018malware}. Unlike traditional detection methods, visualization-based malware analysis generates images from the malware~\cite{ren2020malware}~\cite{vasan2020image}, enabling faster classification than non-visualization methods, as there is no disassembly or execution of applications.
\par In~\cite{tekerek2022novel}, the authors applied the B2IMG algorithm to convert byte files into grayscale and RGB formats. Additionally, they sought to rectify imbalanced data sizes among malware families through a new data augmentation method using CycleGAN. The testing on the BIG2015 and DumpWare10 datasets yielded accuracy rates of 99.86\% and 99.60\%, respectively. The authors in \cite{darem2021visualization} introduce a semi-supervised approach that combines feature engineering, visual analysis techniques, and deep learning to enhance malware detection. The methodology achieves an impressive accuracy rate of 99.12\% in identifying obfuscated malware on the BIG2015 dataset. Researchers in\cite{pinhero2021malware} experimented using grayscale, colored, and Markov images and employed fine-tuned CNN models on the BIG2015 dataset and the Malimg dataset. Gabor images yield F-measures of 99.20\%~(Big2015) and 99.97\%~(MalImg) at 256×256 image dimensions.
\par The FDL-CADIS technique\cite{alzubi2023fusion}, an amalgamation of deep learning models utilizing a MobileNetv2 model to extract features from binary input images followed by hyperparameter tuning using the Black Widow Optimization technique for malware detection. A combination of classifiers, utilizing gated recurrent units (GRUs) and long short-term memory (LSTM), actively analyzes improved representations to identify cyberattacks. Evaluation of benchmark datasets Big 2015 indicates that the proposed approach achieved an F1-score of 98.83\%. The approach in \cite{kalash2018malware} involves converting each malware into images of gray shades and training a CNN for classification. The authors proposed a deep CNN-based architecture (M-CNN) with VGG16 architecture, trained on grayscale images of malware binaries, achieving accuracies of 99.97\% on the BIG 2015 dataset. 
\par Another approach utilizes grayscale images of malware and employs a customized CNN architecture for malware detection, achieving an accuracy of 97.5\% on the BIG2015 dataset, as reported in \cite{gibert2019using}. Gibert et al. present a multimodal approach to malware classification, combining expert-defined features with those acquired through deep learning in \cite{gibert2022fusing}. The method captures N-gram features from assembly language instructions, shapelet-based features, and texture patterns from grayscale images, and the structural entropy of malware. The researchers input these deep features into a gradient-boosting model, utilizing an early-fusion mechanism to combine them with hand-crafted features. The proposed method achieved an accuracy of 99.8\% on the BIG2015 benchmark. Sanjeev et al. introduced DTMIC\cite{kumar2022dtmic} to detect evolving malware threats using transfer learning on the VGG16 architecture. They conducted a comparative analysis with proven CNN architectures, including VGG19, InceptionV3, VGG16, and ResNet50 revealing that DTMIC attains a test accuracy of 93.19\% for Microsoft datasets and remains resilient to packed and encrypted malware. 
\par In \cite{gibert2020hydra}, the authors extracted three modalities: the hex representation and assembly code of malware binary content, the invoked API functions list, and the arrangement of mnemonics in the assembly source code and the byte sequence in the binary content. They achieved 99.51\% F1 score for the BIG2015 dataset. In \cite{naeem2019identification}, the authors actively transformed malware binaries into grayscale images and performed feature extraction and selection on local and global patterns using Gaussian and D-SIFT methods. They also employed methods like hard cluster assignment and regularization to decrease the dimensions of both pattern types. The authors present another approach, MalCVS\cite{xiao2021image}, for classifying malware families by utilizing CoLab which is Colored Label boxes to highlight PE files sections, thereby enhancing section distribution information in the malware image. The approach involves a fine-tuned VGG16 network and a Support Vector Machine as a classifier. Experimental results on the  BIG2015 dataset reveal that MalCVS achieves an F1 score of 97.91\% on the BIG2015 dataset.
\par Jian et al. developed a novel approach, SERLA\cite{jian2021novel}, integrating SEResNet50 (SEnet + ResNet50), Bi-LSTM, and Attention networks. They utilized custom-made three-channel RGB images and implemented data augmentation technology to balance the dataset. The achieved F1-score for the BIG2015 dataset was 98.30\%. 
The authors of~\cite{mallik2022conrec} introduce a Convolutional Recurrence-based technique for malware classification. This method leverages recurrent patterns found in grayscale images of malware samples belonging to the same families. The method involves augmenting the dataset, extracting features with a VGG16-based extractor, and processing through stacked BiLSTM layers, achieving an accuracy of 98.36\% on the BIG2015 dataset. 
\par Currently, researchers are not extensively focusing on integrating images, each crafted from a variety of information sources that could significantly enhance detection performance. We performed an in-depth analysis by developing diverse CNN models, independently applying pre-trained VGG16 to three distinct types of images, and subsequently integrating them using various operators. This entire process delved into detection and classification mechanisms not previously explored in the literature. In \cite{gibert2022fusing}, authors attempted a comparable hybrid model, but it required extensive feature engineering, especially for hand-crafted features. Our deep learning approach completely eliminates the need for such feature engineering, and transfer learning enables precise sample classification. While most researchers primarily focus on refining classifier performance, they notably overlook the crucial dimension of interpretability. This omission is significant because interpretability plays a pivotal role in understanding and trusting the decisions made by machine learning models, particularly in the context of cybersecurity.

\section{Proposed methodology}
\label{sec:proposedmethod}
\begin{figure*}[!htbp]
    \centering
    \includegraphics[scale=0.125]{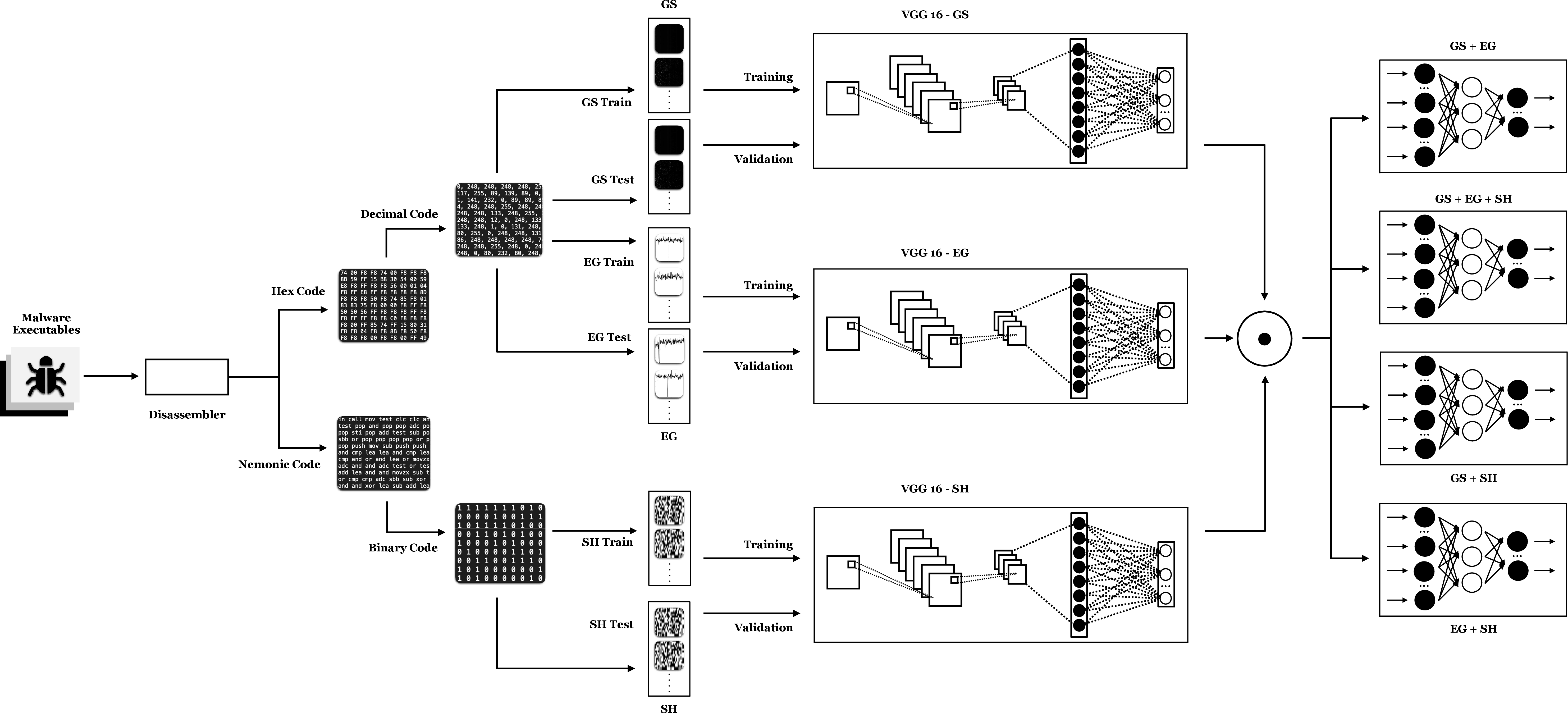}
    \caption{Process of Classification, with three features and VGG16 Models of each feature. Each model is then fused using the operator $\odot$ which can be either \texttt{add, max, avg,} or \texttt{concatenation}}
    \label{fig:process}
\end{figure*}
The proposed malware identification method comprises four primary steps: (1) preprocessing and generating images from various data sources of malware binaries; (2) extracting features using a pre-trained VGG16 model; (3) fusing models using the operator $\odot$ which can be either $add$, $maximum$, $average$, $concatenate$ followed by classifying malware into different families; and (4) interpreting models through the application of gradCAM and t-SNE. Figure \ref{fig:process} illustrates the entire architecture of the proposed malware classification method.

\par

\subsection{Data Collection}

\label{sec:datacollection}
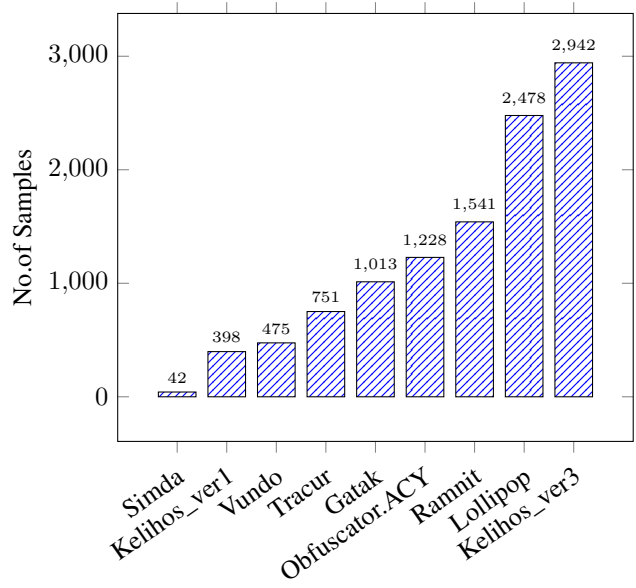
\begin{figure}[!htbp]
    \centering
    \begin{tikzpicture}
        \begin{axis}[
            ybar,                        
            bar width=0.5cm,              
            enlargelimits=0.15,           
            legend style={at={(0.5,-0.15)}, anchor=north,legend columns=-1},
            ylabel={No.of Samples},
            symbolic x coords={Simda, Kelihos\_ver1, Vundo, Tracur, Gatak, Obfuscator.ACY, Ramnit, Lollipop, Kelihos\_ver3},
            xtick=data,
            xticklabel style={rotate=35, anchor=north east}, 
            nodes near coords,
            nodes near coords align={vertical},
            every node near coord/.append style={font=\tiny}
            ]
            \addplot[pattern=north east lines, pattern color=blue] coordinates {
                (Simda,42)
                (Kelihos\_ver1,398)
                (Vundo,475)
                (Tracur,751)
                (Gatak,1013)
                (Obfuscator.ACY,1228)
                (Ramnit,1541)
                (Lollipop,2478)
                (Kelihos\_ver3,2942)
            };
        \end{axis}
    \end{tikzpicture}
    \caption{Highly imbalanced dataset of 9 malware families}
    \label{fig:malware_counts}
\end{figure}

The malware dataset used in this experiment was acquired from the Microsoft Malware Classification Challenge (BIG 2015)~\footnote{Big 2015: https://www.kaggle.com/c/malware-classification}. This dataset emerged as a benchmark in computer security experiments, adopted by numerous researchers for works related to malware detection in various domains~\cite{AHMED202311}~\cite{NI2018871}~\cite{tekerek2022novel}~\cite{kumar2022dtmic}. It comprises 10868 malware samples encompassing a combination of nine families. As illustrated in Figure~\ref{fig:malware_counts}, the distribution of malware samples in Big2015 is highly unbalanced, with a few families having much more than 2000 samples, some ranging from just below 1000 to significantly fewer, as well as more than 1000 obfuscated malware under the obfuscator.ACY family. 




\subsection{Feature Extraction}
\label{sec: featureextracation}
In this work, we focus on three key features: Grayscale (GS) Image, Entropy Graph (EG), and Simhash (SH) Image.\par
\subsubsection{Grayscale Image}
\label{sec: GS}
Malware executables undergo disassembly to generate hexadecimal code which is then converted into decimal. Next, we transform the decimal values in the one-dimensional array into a two-dimensional grayscale image. We use successive pairs of decimal values to increment the array index, determining the pixel intensity of the image, and with repeated values, corresponding indexes will be incremented. Every element in the image array is normalized by dividing them by the sum of its corresponding row and then multiplied by 256 to ensure they are in the range from 0 to 255. 

\par
\subsubsection{Entropy Graph}
\label{sec: EG}
The same decimal codes from which we generated grayscale images are then used to calculate the average entropy and generate the entropy graph. 
To convert decimal to entropy values, we will divide the decimal code into segments, where each segment can have a maximum of 256 decimal elements. Next, the
probability distribution of each element in the segment is calculated and then we compute the entropy by leveraging the Shannon entropy formula.
The Shannon entropy formula is a mathematical equation used to quantify the uncertainty or randomness of a probability distribution. In simpler terms, it measures how much information is contained in malware. The extracted entropy of each segment is divided by the size of the segment to generate average entropy. Next, the average entropy is plotted, where the x-axis ranges from 0 to 256.\par

\subsubsection{Simhash Image}
\label{sec: SH}
\begin{figure}[!htbp]
 \centering
 \includegraphics[scale = 0.3]{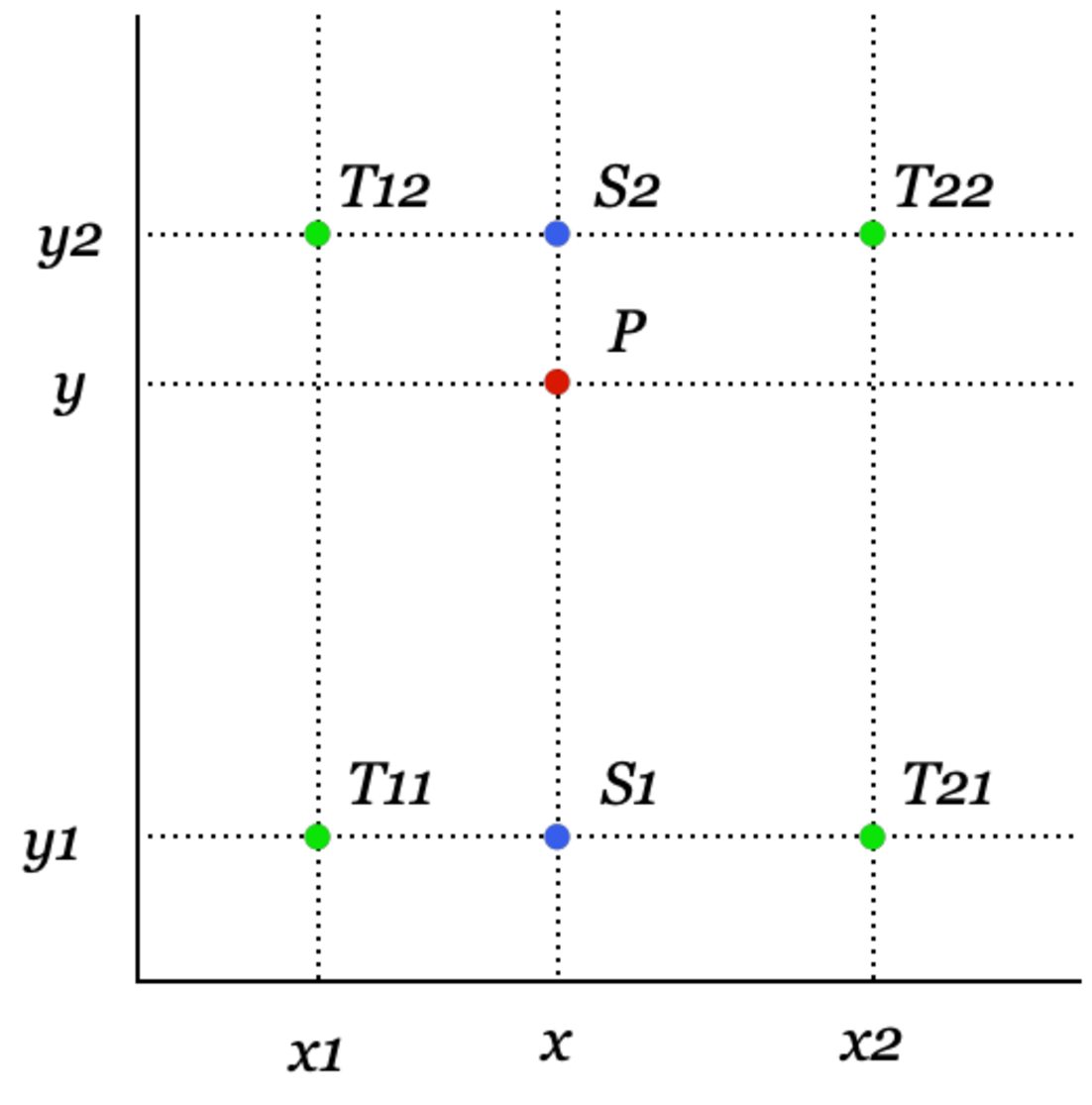}
 \caption{Grid representation of calculating the interpolated pixel \textit{P}. Pixels: \textit{T11, T12, T21, T22} are used used to calculate \textit{$S_{1}$} and \textit{$S_{2}$}, then from which \textit{P}}
 \label{figure:BI}
\end{figure}

\begin{figure}[!htbp]
 \centering
 \includegraphics[width=7cm]{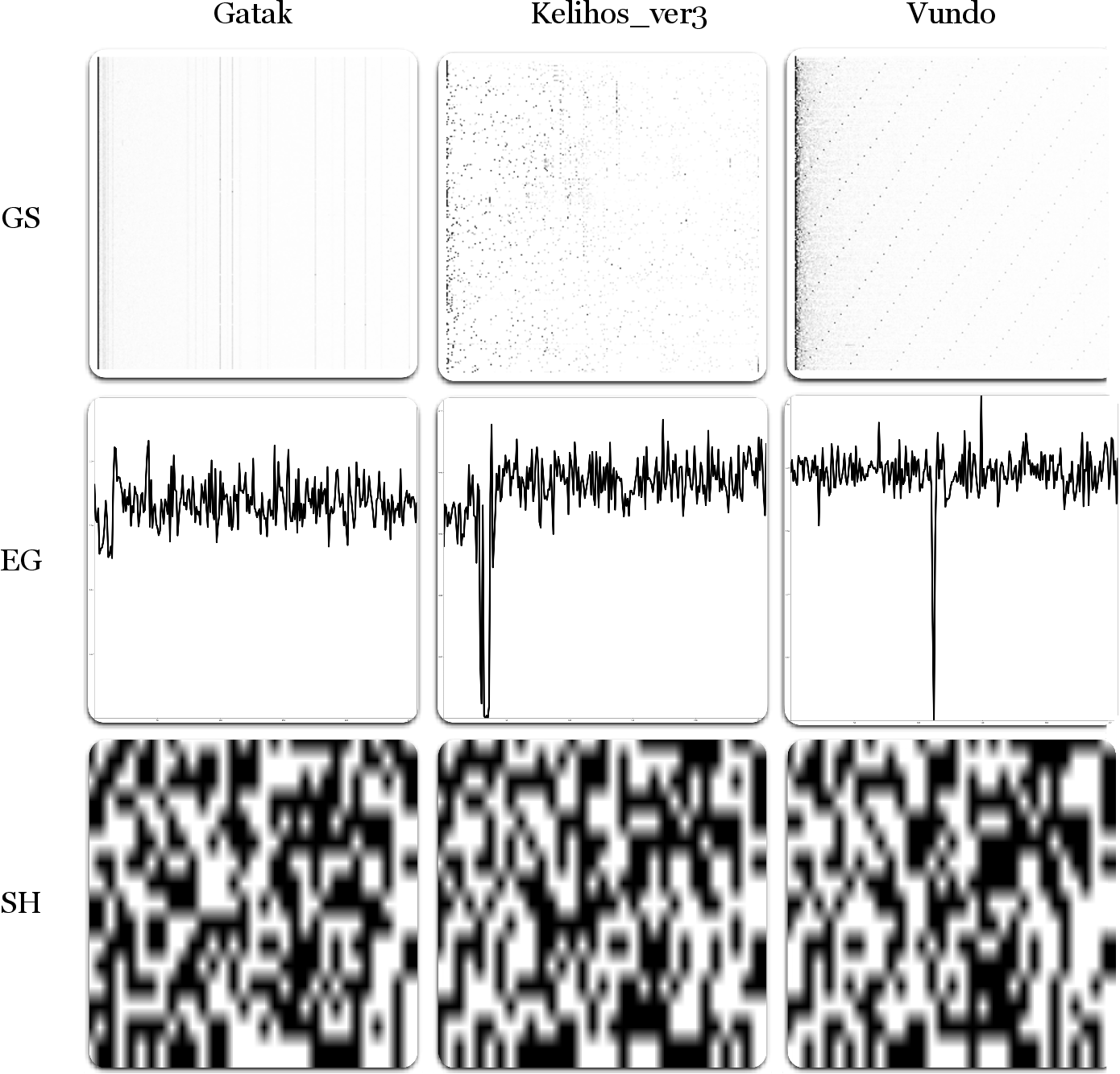}
 \caption{Grayscale, Entropy Graph, and Simhash of malware families (Gatak, Kelihos\_ver3, and Vundo)}
 \label{figure:m_family}
\end{figure}

To generate Simhash feature~\cite{ni2018malware}, we utilize the assembly code from which the mnemonic code or opcodes (e.g., push, mov, call, test, etc.) are extracted. This mnemonic code is then utilized to generate Simhash signatures for each keyword in a given malware sample, using the MD5 hash function. We apply a function $b >> i~\&~1$, where $b$ is the binary representation of the hashed keyword, and $i$ is an iterating variable ranging from 0 to the hash value size. We then use the output of this function to create a binary list, from which we generate a gray-scale image representing the Simhash signature of malware.
To not lose the integrity of the malware by resizing the image and to train the CNN model, we employ bilinear interpolation to zoom the non-square image to a square image of size 224x224. In bilinear interpolation, the original image of size $(m \times n)$ is resized to $(a\times b)$, where $a$ and $b$ are set to 224 in this work, favorable to the VGG architecture. 
When zooming in, each pixel's (\textit{(i, j)}) gray value in the zoomed image correlates to a specific pixel (\textit{$(i \times m/a, j \times n/b)$} in the original image. If the coordinate value is not an integer, we map the gray value to the four nearest pixels in the original image. 
For example, as shown in Figure~\ref{figure:BI}, we use equation~\ref{eq:Px} to calculate the pixel(interpolated point) \textit{P}. The four pixels  around \textit{P}, are \textit{$T_{11}$, $T_{12}$, $T_{21}$, $T_{22}$} which we use to calculate the linear interpolation points \textit{$S_{1}$} and \textit{$S_{2}$} by applying equations~\ref{eq:S1} and~\ref{eq:S2}. We then use \textit{$S_{1}$} and \textit{$S_{2}$} to calculate the bilinear interpolated pixel \textit{P}. The resulting image is illustrated in Figure~\ref{figure:m_family}. The visual differences among the malware family are visible in Figure~\ref{figure:m_family}. A comparison of the entropy graphs for the Gatak, Kelihos\_ver3, and Vundo families reveals their dissimilarity. 

\begin{equation}
\label{eq:S1}
    S_{1}~(x, y)= T_{11}\times((x_{2}-x)/(x_{2}-x_{1})) + T_{21}\times((x-x_{1})/(x_{2}-x_{1}))
\end{equation}
\begin{equation}
\label{eq:S2}
    S_{2}~(x, y)= T_{12}\times((x_{2}-x)/(x_{2}-x_{1})) + T_{22}\times((x-x_{1})/(x_{2}-x_{1}))
\end{equation}
\begin{equation}
\label{eq:Px}
    P(x, y)=S_{1}\times((y_{2}-y)/(y_{2}-y_{1}))+S_{2}\times((y-y_{1})/(y_{2}-y_{1}))
\end{equation}

%

\subsection{Classification Model}
\label{sec: VGG}
\begin{figure}
     \centering
     \includegraphics[scale=.17]{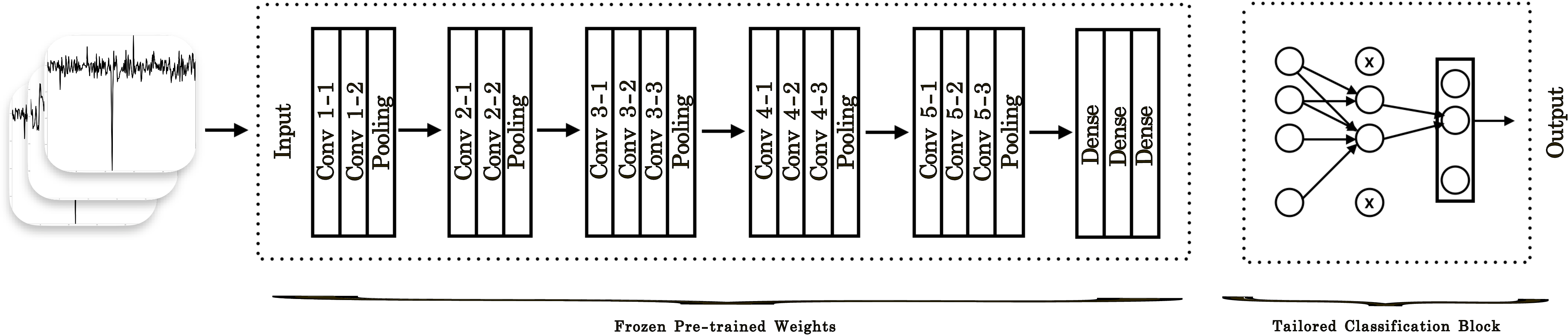}
     \caption{Proposed Architecture}
     \label{fig:vgg16}
 \end{figure}

Our malware classification strategy utilizes an advanced deep neural network architecture incorporating a pre-trained base model, VGG16~\cite{theckedath2020detecting}. VGG16 is a lightweight option among deep convolutional neural networks (CNNs) due to its 16 weight layers. Its ability to learn hierarchical representations helps reveal significant patterns or features in malware images, enhancing malware analysis and detection. This lightweight characteristic of VGG16 circumvents the requirement for substantial computing power~\cite{pinhero2021malware}\cite{rezende2018malicious}. 
The VGG16 architecture consists of 21 layers, including, thirteen convolutional layers, five Max Pooling layers, and three dense layers. However, it uniquely employs only sixteen layers with learnable parameters. Its key innovation is emphasizing 3x3 convolutional filters with stride one and consistently using the same padding alongside 2x2 max pool layers with stride 2. This uniformity extends throughout the architecture, with Conv-1 housing 64 filters, Conv-2 with 128 filters, Conv-3 with 256 filters, and Conv-4 and Conv-5 each accommodating 512 filters. The network also incorporates three Fully-Connected (FC) layers followed by the soft-max layer. Analyzing grayscale images from malware PE files is challenging. However, CNNs like VGG16 utilize pre-trained weights and biases from the ImageNet database, facilitating deployment without requiring dataset-specific retraining. Moreover, notwithstanding the distinctions between natural and malware images, the transference of VGG16 parameters notably enhances malware image recognition through transfer learning. This is primarily due to its proficiency in handling extensive datasets with deeper network layers and smaller filters. The Fig.\ref{fig:vgg16} represents the pre-trained frozen layers and the tailored classification block of vgg-16 used in our proposed method.
\par Our analysis focuses on the performance of VGG-16 models trained on GS, EG, and SH independently. Moreover, we assess how different fusion layers impact the performance of VGG models, employing operations like \texttt{concatenation}, \texttt{add}, \texttt{average}, and \texttt{max} for feature combination. To be specific, we define separate VGG models for independent features, which are then integrated using various merge layers from the Keras library. This process holds significant relevance in diverse neural network architectures, especially those incorporating multiple modalities, potentially enhancing the model's ability to grasp more intricate and meaningful representations.

Specifically, we explore a multi-branch architecture where distinct branches signify separate representations of the input image. The fusion layers take the penultimate layers from multiple independent feature models as input tensors, as illustrated in Figure\ref{fig:process}. Prior research on malware detection has superficially examined model fusion. However, it did not prioritize determining the specific operator to merge different modalities or uncovering the performance gains obtained through fusion.

\par The \texttt{concatenate} function, is specifically leveraged for merging multiple inputs (having the same shape), each containing complementary information, into a single output that carries meaningful representations. We also employ the \texttt{average} function, which condenses multiple input tensors into a single tensor by computing their element-wise averages, aiding in building more generalized representations, promoting translation invariance, and supporting the creation of compact final classification layers. Further, we utilize the addition function (\texttt{add()}) in deep neural networks that add corresponding elements of two tensors, creating a new tensor. It merges output features from different models, enriching representations of input malware images and leveraging model strengths for improved accuracy and generalization. Furthermore, we used the maximum function (\texttt{max()}), essential for selecting the maximum element-wise value between two tensors. It serves a crucial role in combining features, enhancing representations, and introducing non-linearity. 

\par
\begin{figure*}[!htbp]
    \centering
    \includegraphics[scale=0.42]{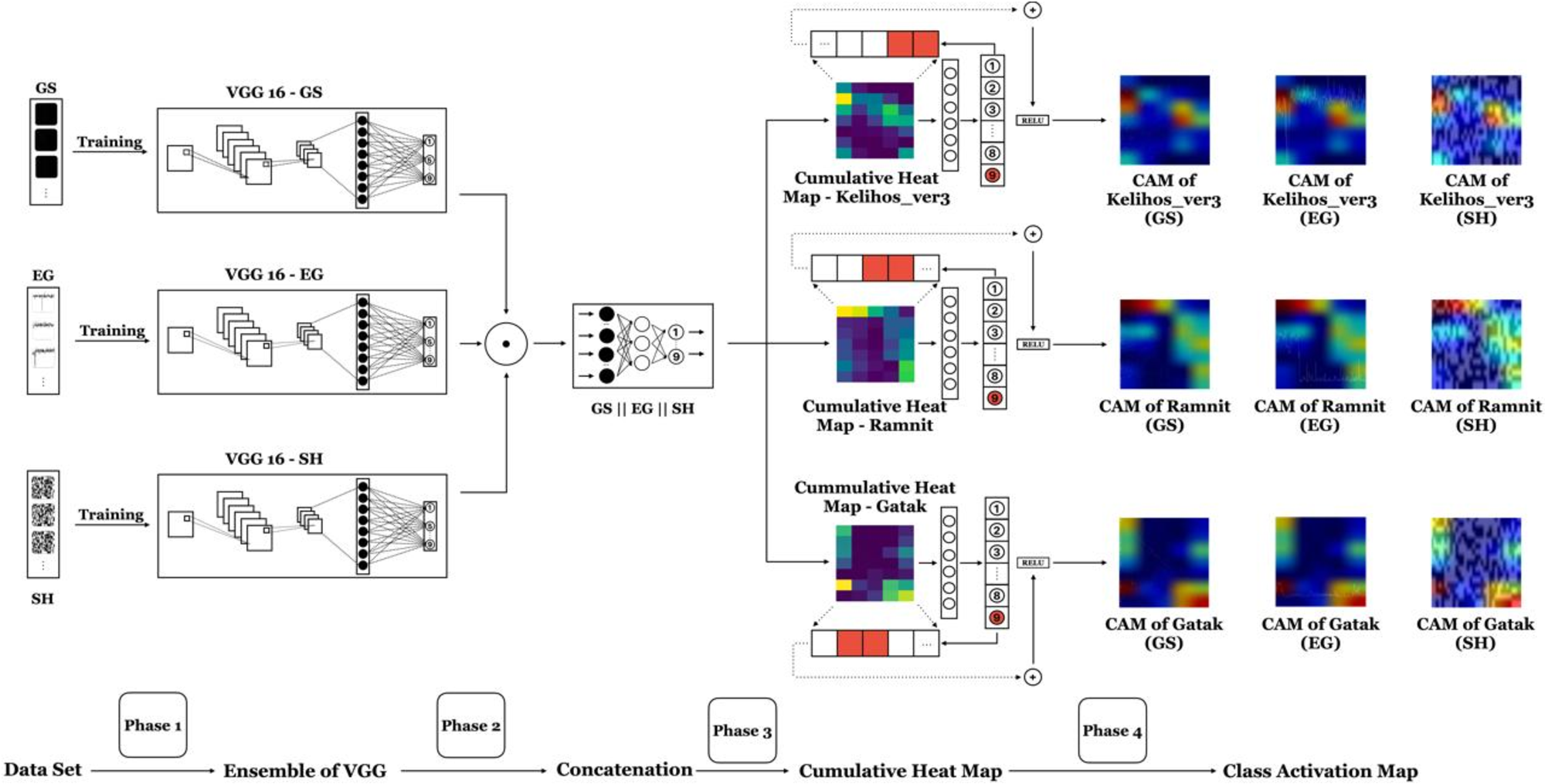}
    \caption{The creation of Grad-CAM includes the feature extraction process using the VGG16 network and three fully connected layers as classifiers.}
    \label{fig:vgg16gradcam}
    
\end{figure*}

 
Anti-malware systems developed using traditional machine learning algorithms often result in opaque models, leaving security analysts unable to discern the features influencing the classifier's decisions. Our approach enhances model interpretability by leveraging Gradient Class Activation Map (Grad-CAM)~\cite{selvaraju2017grad}, empowering malware analysts to provide valuable feedback to classification models~(refer to Figure~\ref{fig:vgg16gradcam}). In the context of malware classification, explainable VGG-16 models, with the aid of class activation maps, identify regions of features in malicious samples that contribute most to the categorization. This information supports analysts in understanding the characteristics of malware, including the prominent byte codes learned by the classifier.

\section{Experiments and Results}
\label{sec:experimentsandresults}
 \subsection{Experimental Set-up}
 \label{sec:expenv}
This paper utilizes a system powered by MacOS 14.2.1 with an M1 Chip, 8GB RAM, and 128GB of storage for the conversion of hex codes of malware into hex images, entropy graphs, and simhash images. We employed a Colab Pro+ setup, which included 40GB GPU RAM, 83.5GB of system RAM, and 166.8GB of storage, for training and testing independent feature VGG models as well as combined features VGG models. For our experiments, we chose the extensively utilized BIG2015~\cite{zhu2021malware} dataset as mentioned in the section \ref{sec:datacollection}. In the raw data, each file is represented using a hexadecimal, showing its binary content. The header, which is the initial part of the file, is deliberately left out to maintain sterility. To minimize the impact of random error, each experiment was repeated five times and the averaged results were reported. The following experiments were performed using different visual malware features:

 \begin{itemize}
     \item \textit{\textbf{Experiment-1:}} Effectiveness of GS, EG, and SH VGG16 models in classifying malware binaries (\textit{answers $RQ_1$}).
     
     \item \textit{\textbf{Experiment-2:}} Assessing the merging of malware visualization models using various merge operators (\textit{answers $RQ_2$}).
     
     \item \textit{\textbf{Experiment-3:}} 
     Interpretability of the proposed method (\textit{answers $RQ_3$}).
 \end{itemize}
\subsection{Evaluation Metrics}
\label{sec: evaluationmetrics}
This work employs evaluation metrics such as Accuracy, Precision, and F1 Score. Accuracy measures the model's correct prediction frequency, Precision quantifies correctly predicted positives among all predicted positives, Recall reflects the ratio of True Positives to actual positives, and F1-score balances Precision and Recall's performance. Equations~\ref{eq:A} to equation~\ref{eq:F1} illustrate the diverse metrics applied in this study. Given the imbalanced nature of the dataset, we prioritize the F1-score over other metrics to assess the model's performance. Besides the above-mentioned metrics, we also estimate the time required to predict a file using our proposed models.
\begin{equation}
\label{eq:A}
    Accuracy~(A)= (TP + TN)/(TP + FP + TN + FN )
\end{equation}
\begin{equation}
\label{eq:P}
    Precision~(P) = TP/(TP + FP)
\end{equation}
\begin{equation}
\label{eq:R}
    Recall~(R) = TP/(TP + FN)
\end{equation}
\begin{equation}
\label{eq:F1}
    F1-measure~(F1) = 2\times (P\times R)/(P + R)
\end{equation}

\subsection{\textbf{Experiment-1:} Effectiveness of GS, EG, and SH VGG16 models in classifying malware binaries}
\label{sec:experiment-1}
Initially, we perform experiments to assess the performance of three separate VGG16 models. These models are trained independently on grayscale, entropy graph, and simhash images, respectively. Table~\ref{tab:single} presents the comprehensive classification outcomes for these VGG16 models. Notably, the models trained on GS and EG images achieved high F1-scores of 0.997 and 0.998, respectively. However, the simhash-based VGG16 model showed a decrease in its F1-score. These results also demonstrate uniform prediction times (0.01 seconds) for a single instance across all models.

\begin{table} [!htbp]
\centering
 \caption {VGG16 Independent Feature Performance Comparison using Accuracy, Precision, Recall, F1-Measure, and Prediction Time(in seconds)}
 \label{tab:single}
  \begin{tabular}{  c  c  c  c  c  c }
    \hline
    \textit{\textbf{Feature}} & \textit{\textbf{A}} & \textit{\textbf{P}} & \textit{\textbf{R}} & \textit{\textbf{F1}} & \textit{\textbf{Time}} \\\midrule
    $GS$ & 0.998 & 1.0 & 0.996 & 0.997 & 0.01\\ 
    $EG$ & 0.999 & 0.998 & 0.998 & 0.998 & 0.01\\
    $SH$ & 0.996 & 0.966 & 0.963 & 0.963 & 0.01 \\ \bottomrule
  \end{tabular}
\end{table}

\begin{center}
\fcolorbox{black}{lightgray}{
\begin{minipage}[center]{0.97\linewidth}
\textbf{$RQ_1$ Summary:} \em{Training VGG16 models on grayscale and entropy graph images yields high classification performance, while using simhash images results in decreased F1 and detection rate, emphasizing the influence of image type on classification performance and the need for further investigation.} 
\end{minipage}}
\end{center}
\subsection{\textbf{Experiment-2:} Assessing the merging of malware visualization models using various merge operators}
\label{sec:experiment-2}

We proceed to evaluate the effectiveness of both \textit{bi-modal} and \textit{multimodal} representations of a given malware binary in our classification task. These features aim to capture complementary or distinct information from each modality, which can then be used for classification. To achieve this, we have developed a VGG16 model that trains on the fusion of feature representations from multiple modalities. Our objective is to investigate whether combining information from different modalities enhances the model's understanding and performance. Specifically, we study the influence of four fusion operators ($\odot$, refer Figure~\ref{fig:process}), namely \texttt{max}, \texttt{avg}, \texttt{add}, and \texttt{concatenate}, in identifying malware samples. Each of these operators combines two tensors representing the most significant features extracted from individual modalities. A typical bi-modal representation of malware images involves combining two independent modalities, such as (GS$\odot$EG), (GS$\odot$SH), or (EG$\odot$SH). However, multimodal representation incorporates all three image representations (GS$\odot$EG$\odot$SH) simultaneously for training and predicting executable images.
\par Table~\ref{tab:max} illustrates the outcomes achieved by employing various modalities with the \texttt{max} merging operator. The combination of two modalities resulted in F1-scores and detection rates ranging from 0.922 to 0.974 for the VGG16 model. Notably, merging features from all three modalities yielded an F1-score of 0.985. However, when compared to independent models, except for the SH VGG16 model (refer Table~\ref{tab:single}), bi-modal VGG16 models demonstrated lower F1-scores. It's essential to highlight that the merged classifier encounters challenges in effectively extracting meaningful insights from the element-wise maximum presentations of tensors for classifying malware families.
\begin{table} [!htbp]
\centering
 \caption {Feature Model Performance metrics and Prediction Time (in seconds) with \texttt{max} fusion layer}
 \label{tab:max}
  \begin{tabular}{ l l  l l  l l }
    \hline
    \textit{\textbf{Feature}} & \textit{\textbf{A}} & \textit{\textbf{P}} & \textit{\textbf{R}} & \textit{\textbf{F1}} & \textit{\textbf{Time}} \\ \midrule
    $max(EG,SH)$ & 0.922 & 0.927 & 0.922 & 0.922 & 0.090 \\
    $max(GS,SH)$ & 0.959 & 0.962 & 0.959 & 0.957 & 0.090 \\ 
    $max(GS,EG)$ & 0.974 & 0.975 & 0.974 & 0.974 & 0.097 \\ 
    $max(GS,EG,SH)$ & 0.986 & 0.985 & 0.986 & 0.985 & 0.136 \\ \bottomrule
  \end{tabular}
\end{table}

\par Analyzing the performance of merged VGG16 models using bi-modal and multimodal image representations, as shown in Table~\ref{tab:avg}, provides significant insights. When considering all three modalities simultaneously, we observe an F1-score of 0.985, consistent with the previous scenario (refer to Table~\ref{tab:max}). Comparing the \texttt{max} fusion operator reveals nearly identical trends in detection rates, F1-score, and prediction time. It is worth mentioning that the merging of feature representations of simhash images to create a bi-modal feature representation degrades the performance of the classifier in comparison to its counterparts. The primary drawback of averaging is that it condenses multiple features into a single value, eliminating crucial details that are unique to each feature. When averaging the two feature representations of a sample, we may forfeit the distinct characteristics of each representative modality.
\begin{table} [!htbp]
\centering
 \caption {Performance and Prediction Time (in seconds) of feature models with \texttt{avg} fusion layer}
 \label{tab:avg}
  \begin{tabular}{ l  l  l  l  l  l  l}
    \hline
    \textit{\textbf{Feature}} & \textit{\textbf{A}} & \textit{\textbf{P}} & \textit{\textbf{R}} & \textit{\textbf{F1}} & \textit{\textbf{Time}} \\ \midrule
    $avg(EG,SH)$ & 0.934 & 0.932 & 0.934 & 0.932 & 0.088 \\ 
    $avg(GS,SH)$ & 0.936 & 0.948 & 0.936 & 0.937 & 0.093 \\ 
    $avg(GS,EG)$ & 0.974 & 0.975 & 0.974 & 0.973 & 0.091 \\
    $avg(GS,EG,SH)$ & 0.986 & 0.985 & 0.986 & 0.985 & 0.136 \\ \bottomrule
  \end{tabular}
\end{table}
\par Furthermore, we analyzed the performance of the merged VGG16 classifier utilizing the \texttt{add} operator. Table~\ref{tab:add} displays a slight improvement in the F1-score compared to the bi-modal deep learning models. The prediction performance of the classifier trained using multiple modalities follows the trends observed in the previous two scenarios (refer to Table~\ref{tab:max} and Table~\ref{tab:avg}). This enhanced performance is primarily attributed to the retention of information from distinct tensors when combining them using addition, contrasting with the potential diminishment of individual element significance in averaging or maximum operations. Addition assigns equal importance to corresponding elements, potentially amplifying the impact of significant features. Furthermore, addition can effectively handle noise or outliers present in the tensors, preserving their influence, while averaging might dilute their impact.
\begin{table} [!htbp]
\centering
 \caption {Evaluation Metrics values and Prediction Time (in seconds) of features models with \texttt{add} fusion layer}
 \label{tab:add}
  \begin{tabular}{ l  l  l  l  l  l  l}
    \hline
    \textit{\textbf{Feature}} & \textit{\textbf{A}} & \textit{\textbf{P}} & \textit{\textbf{R}} & \textit{\textbf{F1}} & \textit{\textbf{Time}} \\ \midrule
    $EG+SH$ & 0.936 & 0.932 & 0.936 & 0.934 & 0.090 \\ 
    $GS+SH$ & 0.968 & 0.970 & 0.968 & 0.969 & 0.091 \\ 
    $GS+EG$ & 0.987 & 0.981 & 0.981 & 0.981 & 0.090 \\ 
    $GS+EG+SH$ & 0.983 & 0.983 & 0.983 & 0.983 & 0.137 \\ \bottomrule

  \end{tabular}
\end{table}
\par
\begin{figure}[!htbp]
    \centering
    \includegraphics[scale = 0.5]{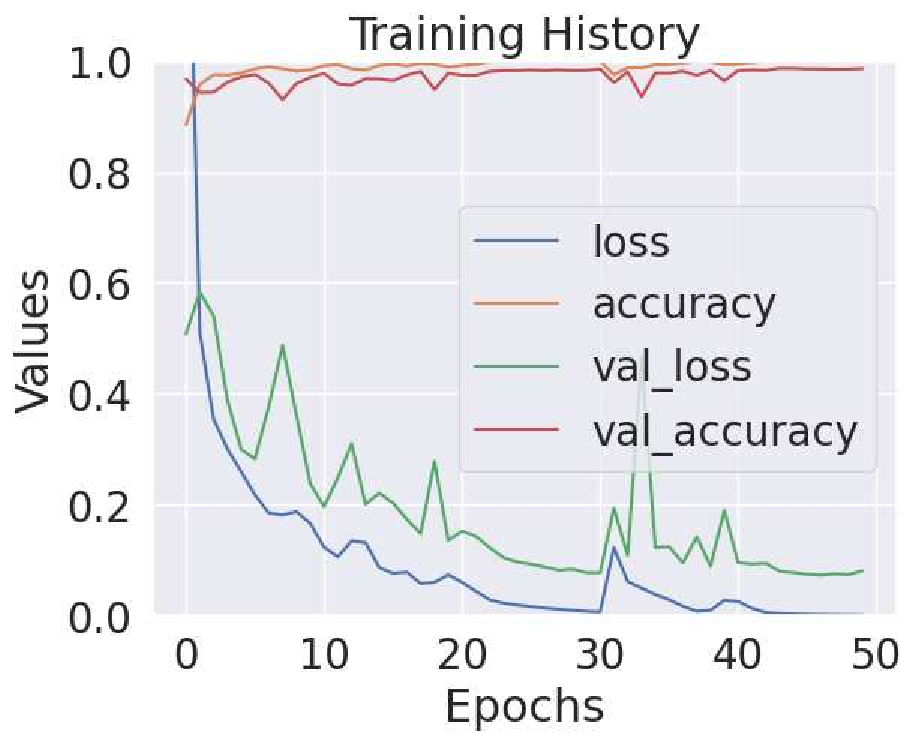}
    \caption{Training History : $GS||EG||SH$}
    \label{fig:thCONCAT}
\end{figure}

 Finally, we assessed the performance of merged VGG16 models by employing the concatenation operation on tensors, as demonstrated in Table~\ref{tab:concat}, which consolidates feature representations from distinct modalities associated with malware images. Our observations indicate a significant enhancement in the classifier's performance (F1-score and the detection rates in the range of 0.999 to 1.0) compared to all independent VGG16 models (Table~\ref{tab:single}). Moreover, the F1-score of both bi-modal and multimodal concatenation CNN models surpasses that of alternative fusion operators such as \texttt{max}, \texttt{avg}, and \texttt{add}. An essential aspect of the concatenation operator is its ability to preserve feature representations from each modality within the merged tensor, resulting in a richer representation that captures diverse information from grayscale, entropy graph, and simhash modalities. Additionally, concatenation augments parameters and enhances the model's capacity to learn intricate representations from the comprehensive feature space of different modalities. This approach enables a more nuanced and comprehensive representation of multimodal data, potentially enhancing the model's performance in tasks requiring multimodal fusion. The training and validation curve is depicted in Figure~\ref{fig:thCONCAT}.

\begin{table} [!htbp]
\centering
 \caption {VGG16: Accuracy, Precision, Recall, F1, \& Prediction Time (in seconds) of feature models with \texttt{concatenate} fusion layer.}
 \label{tab:concat}
  \begin{tabular}{ l  l  l  l  l  l}
    \hline
    \textit{\textbf{Feature}} & \textit{\textbf{A}} & \textit{\textbf{P}} & \textit{\textbf{R}} & \textit{\textbf{F1}} & \textit{\textbf{Time}} \\ \midrule
    $EG||SH$ & 0.999 & 0.999 & 0.999 & 0.999 & 0.095 \\ 
    $GS||SH$ & 1.0 & 1.0 & 1.0 & 1.0 & 0.094 \\ 
    $GS||EG$ & 1.0 & 1.0 & 1.0 & 1.0 & 0.093 \\ 
    $GS||EG||SH$ & 1.0 & 1.0 & 1.0 & 1.0 & 0.139 \\ \bottomrule

  \end{tabular}
\end{table}

\par \textbf{Family-wise classification:} Table~\ref{tab:familyclassification} showcases the classification of various families by independent, bi-modal, and multimodal classifiers. Each family, represented as $Fi$ in the BIG2015 malware dataset, corresponds to values indicating the detection rate. A closer examination through vertical scanning of the table highlights superior classifier performance when values approach one in multiple entries. Specifically, the VGG16 model, trained using both grayscale and entropy graph data, effectively distinguishes between two distinct malware families. The classifier, trained on grayscale image patterns, achieved a perfect 100\% detection rate for families $F3$ and $F4$. Likewise, the model trained using entropy graph data demonstrated comparable detection rates for classes $F7$ and $F9$. Furthermore, the simhash model exhibited a detection rate exceeding 0.95 for eight families and successfully identified 78.3\% of malware variants associated with family $F5$.

\par We devised twelve bi-modal VGG16 classifiers, each utilizing different fusion techniques. Among these, the three models employing concatenation achieved 1.00~(100\%) detection across all malware families. In contrast, the remaining eight VGG16 models, using merging operators such as \texttt{max}, \texttt{avg}, and \texttt{add}, fell short of achieving perfect (100\%) detection rates for any malware family. Particularly, these models displayed inadequate performance, ranging from 0.00~(0\%) to 0.625~(62.5\%) in identifying family $F7$. Similar trends persisted across multiple modalities (GH, EG, and SH) for these operators, with detection rates ranging from 0.250 to 0.437 in identifying family $F7$. Conversely, when incorporating the feature representations of GH, EG, and SH through concatenation, the VGG16 model achieved perfect (100\%) detection rates across all families. 
\par \textbf{Detection of obfuscated binaries:} To assess the effectiveness of our multimodal models in detecting obfuscated malware, we evaluated their ability to identify PE files belonging to the \texttt{Obfuscator.ACY} family~(F5 in Table~\ref{tab:familyclassification}). This class of malware employs sophisticated techniques to evade detection by anti-malware systems, and Windows Defender Antivirus~\footnote{Obfuscator.ACY: https://www.microsoft.com/en-us/wdsi/threats/malware-encyclopedia-description?Name=VirTool:Win32/Obfuscator.ACY} has categorized these variants as high-alert threats. The detection of \texttt{Obfuscator.ACY} binaries, which are particularly challenging due to their obfuscation techniques, provide valuable insights into the robustness of our proposed models.

\par Table~\ref{tab:familyclassification} presents the detection rates for individual malware families. We observe that models employing independent features (GS, EG, or SH) achieve a detection rate ranging from 0.783 to 0.979, with the VGG16 classifier trained on SH images exhibiting suboptimal performance. Additionally, we note that classification models trained by fusing two or three modalities using operators like \texttt{avg}, \texttt{max}, and \texttt{add} achieved detection rates between 0.885 and 0.989, particularly when SH image representations were combined with GS and EG. Notably, the models trained with the concatenation of multiple image feature representations (two or three) demonstrated the ability to accurately identify obfuscated samples in the test set~(detection rate between 0.998 to 1.0). This underscores the effectiveness of multimodal approaches, especially using concatenation operators in combating obfuscated malware.

\begin{center}
\fcolorbox{black}{lightgray}{
\begin{minipage}[center]{0.97\linewidth}
\textbf{$RQ_2$ Summary:} \em{Concatenation emerges as a robust approach for multimodal data fusion, facilitating effective detection of sophisticated malware threats, including obfuscated binaries, across diverse malware families.} 
\end{minipage}}
\end{center}

\subsection{\textbf{Experiment-3:} Interpretability of the proposed method }
\begin{figure}
  \begin{center}
    \includegraphics[scale=0.3]{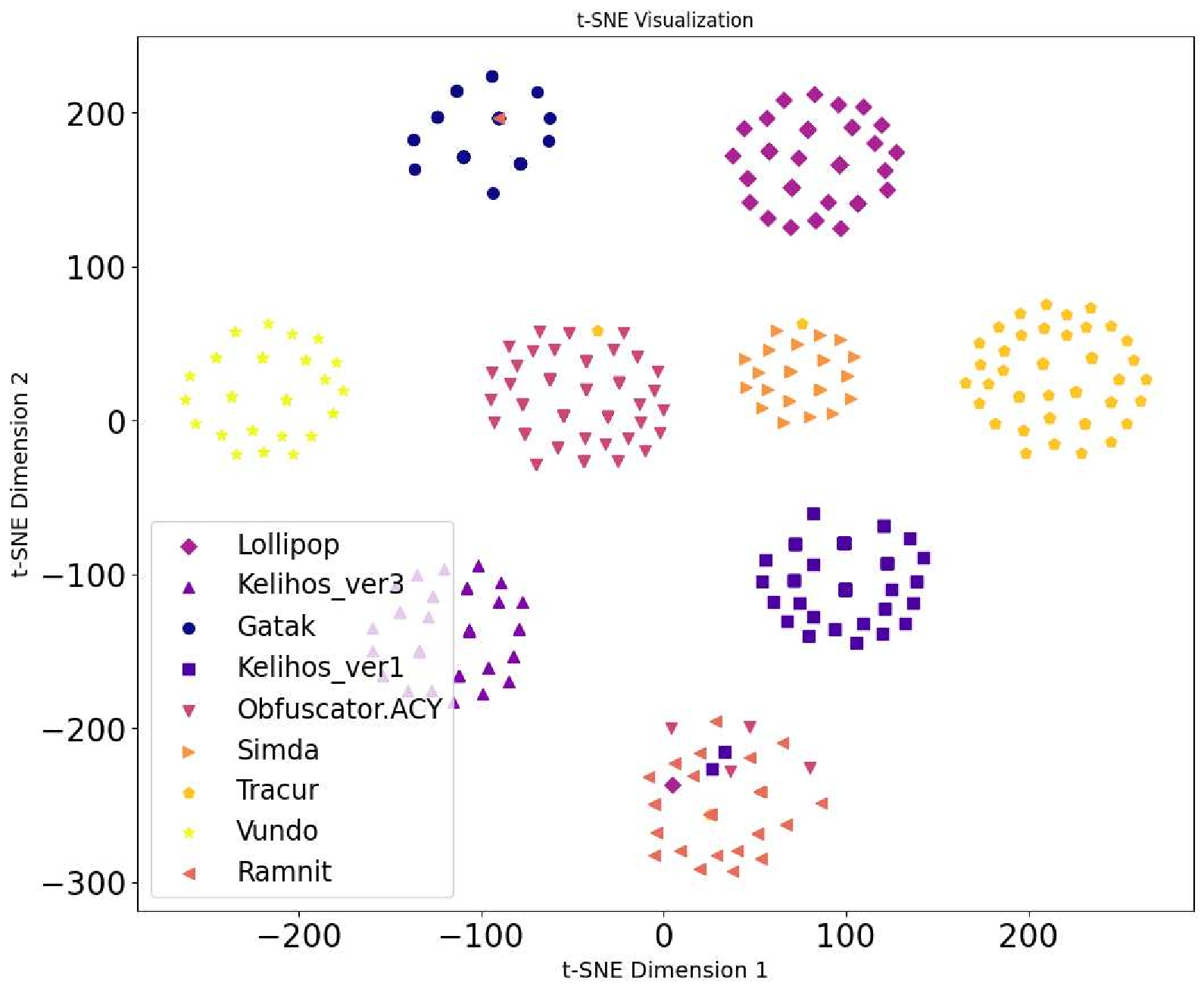}
  \end{center}
  \caption{t-SNE $GS||EG||SH$}
  \label{fig:t-SNE-Concatmodel}
\end{figure}
To demonstrate the potential of using models trained on multiple data sources, we further analyzed the features extracted from such models. In particular, we sought to enhance the interpretability of the outcomes by visualizing the results obtained using different estimators. This visualization would increase the trust of security personnel in adopting the developed models for future use in malware classification. To improve the interpretability of the results, we created t-distributed stochastic neighbor embedding~(t-SNE) plots for different merged models, particularly focusing on the outcomes of multimodal image representations, which exhibited superior performance compared to bi-modal cases.
\par
\begin{figure}
    \centering
    \begin{subfigure}[b]{0.31\textwidth}
        \includegraphics[width=1\textwidth]{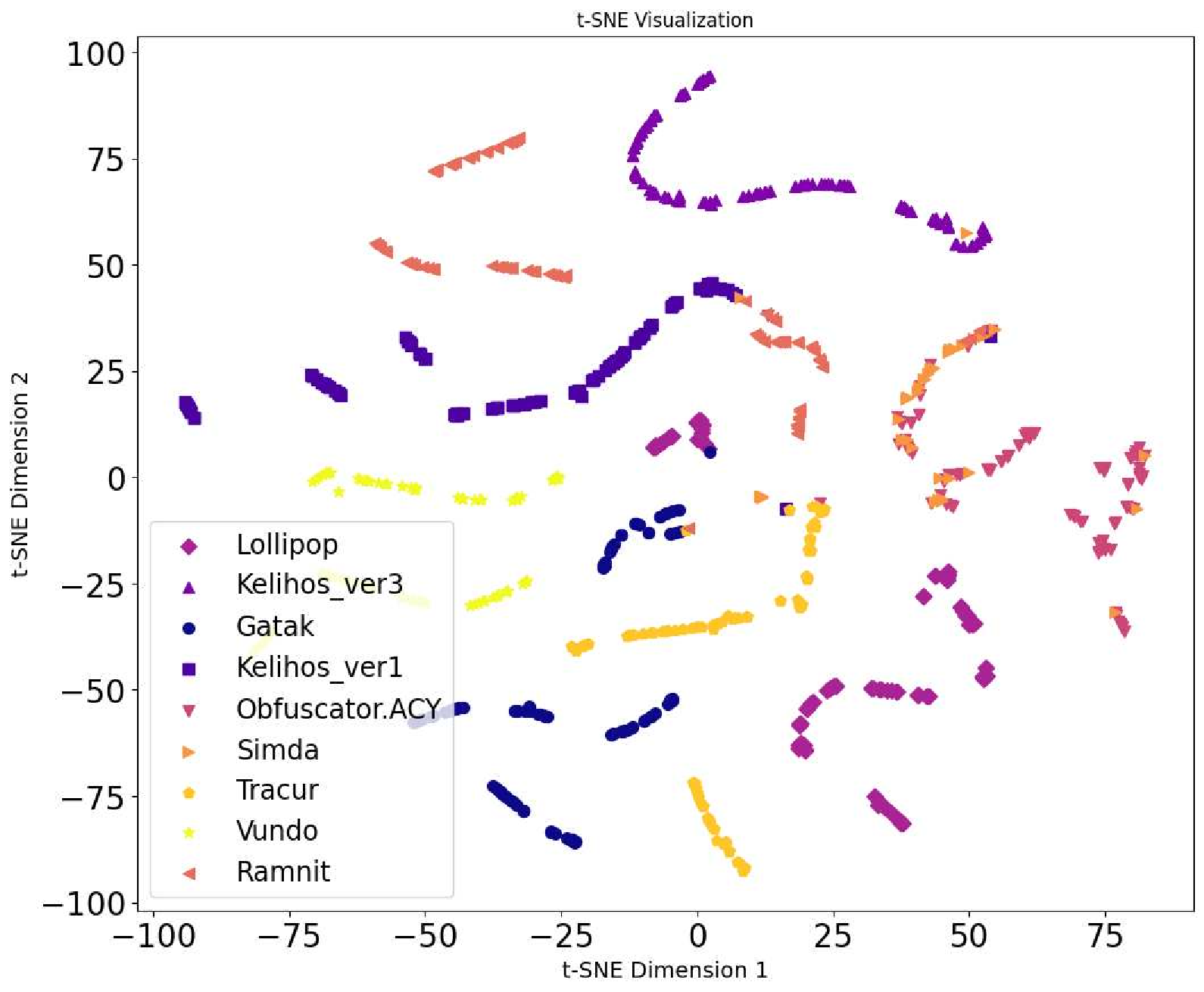}
        \caption{$max(GS,EG,SH)$}
        \label{fig:tsneCONCATuni}
    \end{subfigure}
    ~ 
    \begin{subfigure}[b]{0.31\textwidth}
        \includegraphics[width=1\textwidth]{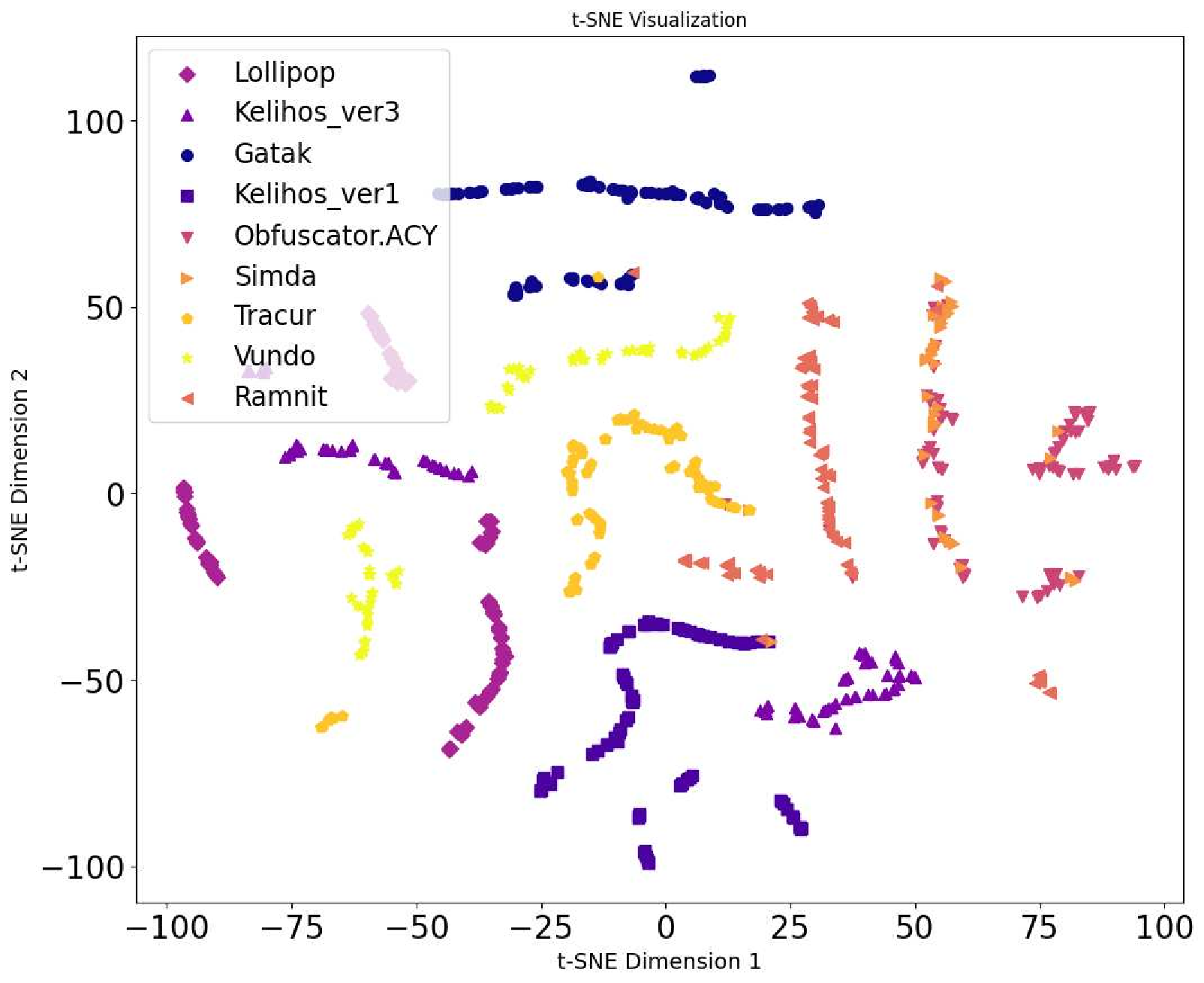}
        \caption{$avg(GS,EG,SH)$}
        \label{fig:tsneAVG}
    \end{subfigure}
    ~ 
    \begin{subfigure}[b]{0.31\textwidth}
        \includegraphics[width=1\textwidth]{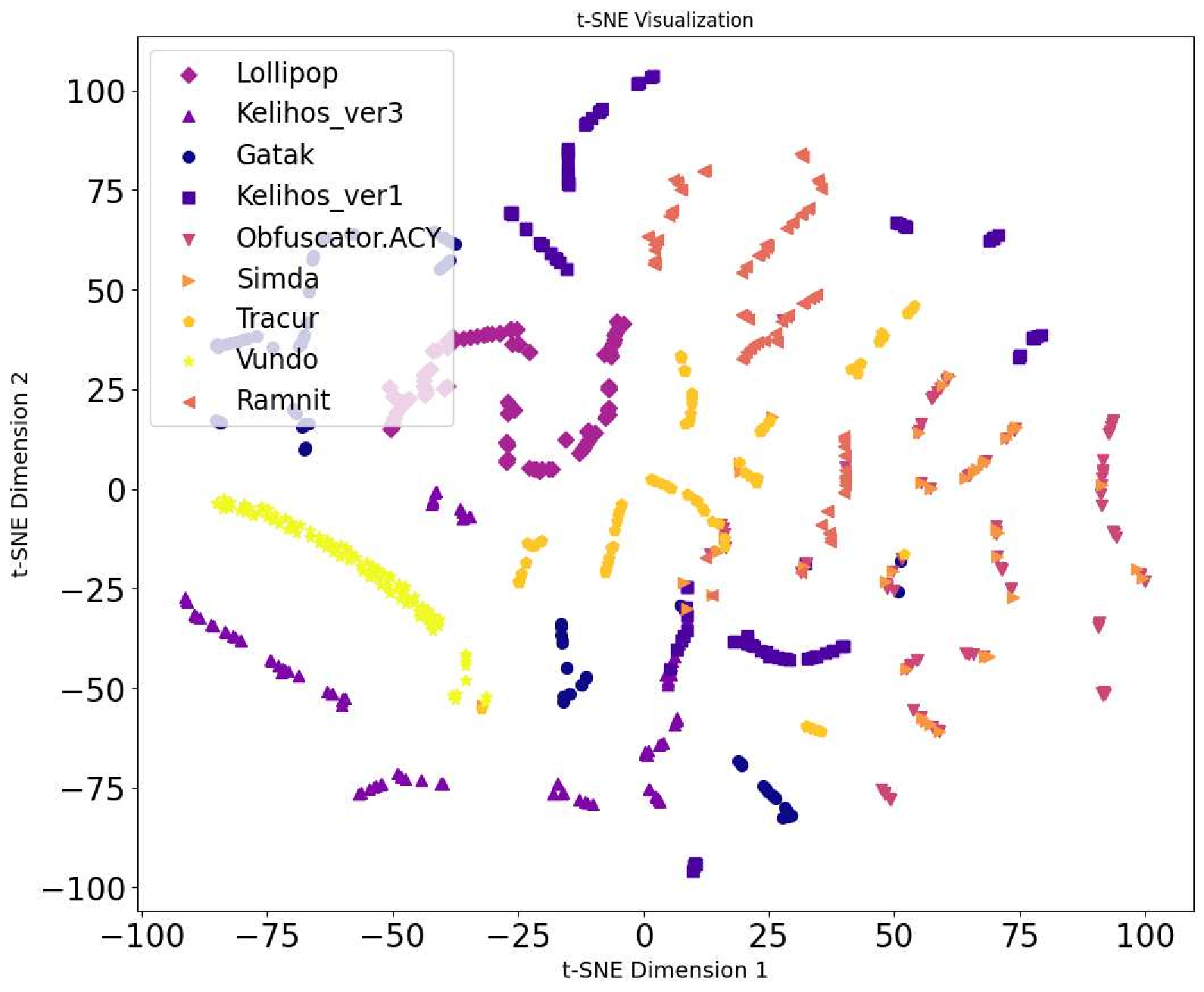}
        \caption{$add(GS,EG,SH)$}
        \label{fig:tsneADD}
    \end{subfigure}
    \caption{tSNE Visualisation}\label{fig:tSNE}
\end{figure}

 We employed t-SNE, a powerful dimensionality reduction technique. t-SNE transforms the multi-dimensional feature space into a comprehensible two-dimensional representation while preserving the underlying relationships between data points. In our study, we specifically focused on the feature extractor layer, a crucial intermediary layer preceding the prediction layer in merged models. By leveraging t-SNE, we successfully visualized the extracted features, demonstrating the effectiveness of our approach. Figures~\ref{fig:t-SNE-Concatmodel} to \ref{fig:tsneADD} showcase the visualizations of the four fusion VGG16 models. As observed, the data points are grouped, and the families are distinctly distinguishable in the \texttt{concatenation} model (refer Figure~\ref{fig:t-SNE-Concatmodel}) compared to other VGG16 models (cases where scattered data points are observed, there is also overlapping clusters, refer to Figures~\ref{fig:tsneCONCATuni} to \ref{fig:tsneADD}).

\par In addition, to visualize the higher-level features extracted by VGG16, we employ Grad-CAM~\cite{selvaraju2017grad}. VGG16's convolutional layers capture both high-level semantics and spatial information from malware images. However, these features become less apparent when passed through dense layers for classification. To address this, we leverage the last convolutional layer and implement Grad-CAM, a technique that analyzes the gradient flow of each neuron in this layer to highlight the most informative regions for class prediction. To visualize prominent byte patterns for distinct malware families, we developed cumulative class activation maps for malware instances corresponding to each family. Aggregating class activation maps (CAMs) across multiple instances yields a cumulative representation of feature importance. The highlighted regions in CAMs depict areas that influence the model's predictions across families. By examining these activated regions, malware analysts can identify the specific features that the model focuses on and understand how these features contribute to the classification outcome. Also, the prominent regions in the activation maps can be used to develop family signatures, which can be used to categorize malware instances. 
\par To grasp visual trends for each family, we generated a cumulative heatmap through the following steps: (i) initiated the creation of an empty heatmap with the desired dimension, (ii) incorporated data points from the heatmaps of each malware sample, (iii) normalized the heatmap to ensure all values within its range fall within the desired bounds, and (iv) applied a color map to the normalized heatmap, resulting in a cumulative class activation map of the desired size. 
\par
\begin{figure}
    \centering
    \begin{subfigure}[b]{0.31\textwidth}
        \includegraphics[width=1\textwidth]{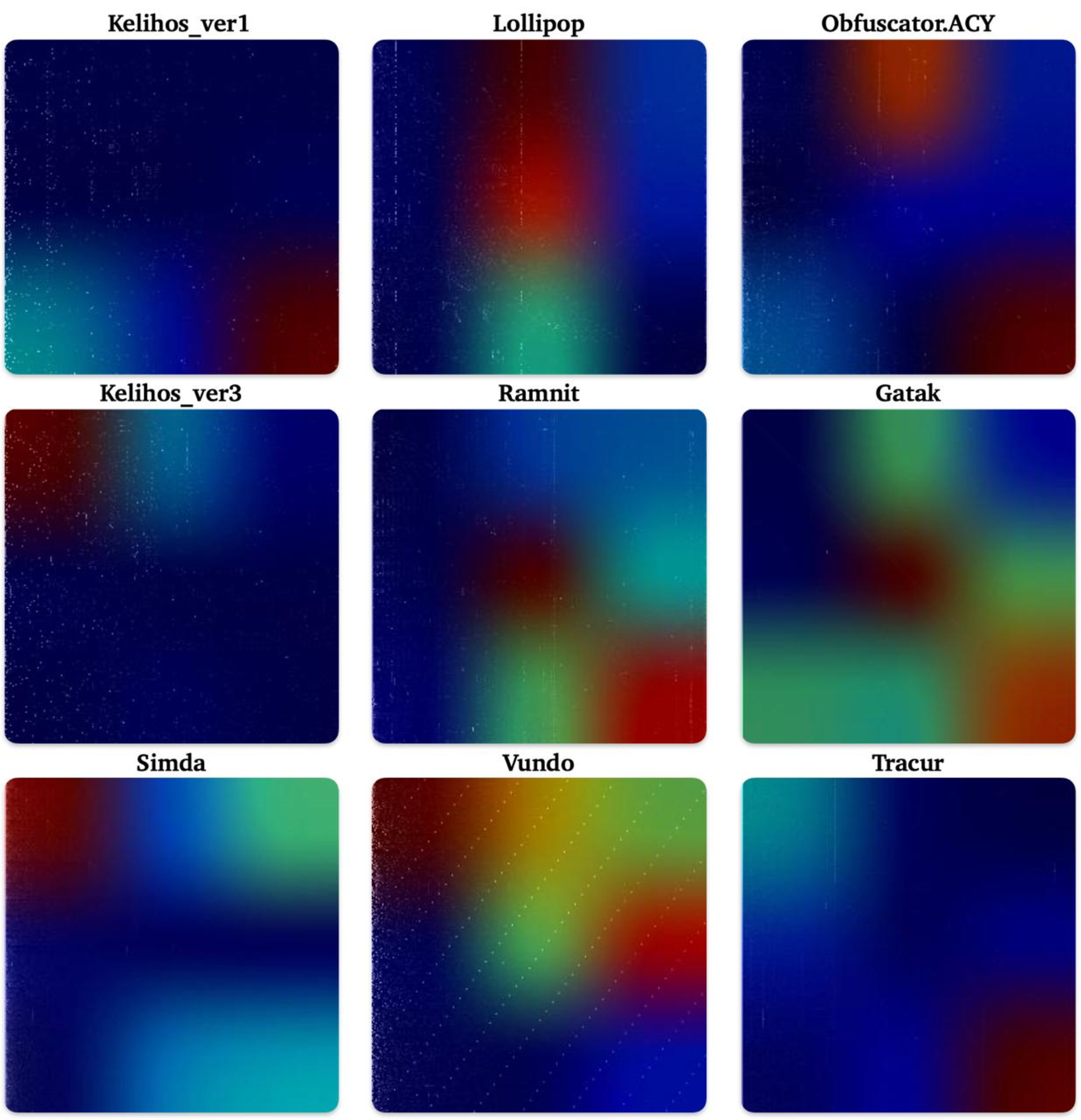}
        \caption{$GS$}
        \label{fig:gradCAM[GS]}
    \end{subfigure}
    ~ 
    \begin{subfigure}[b]{0.305\textwidth}
        \includegraphics[width=1\textwidth]{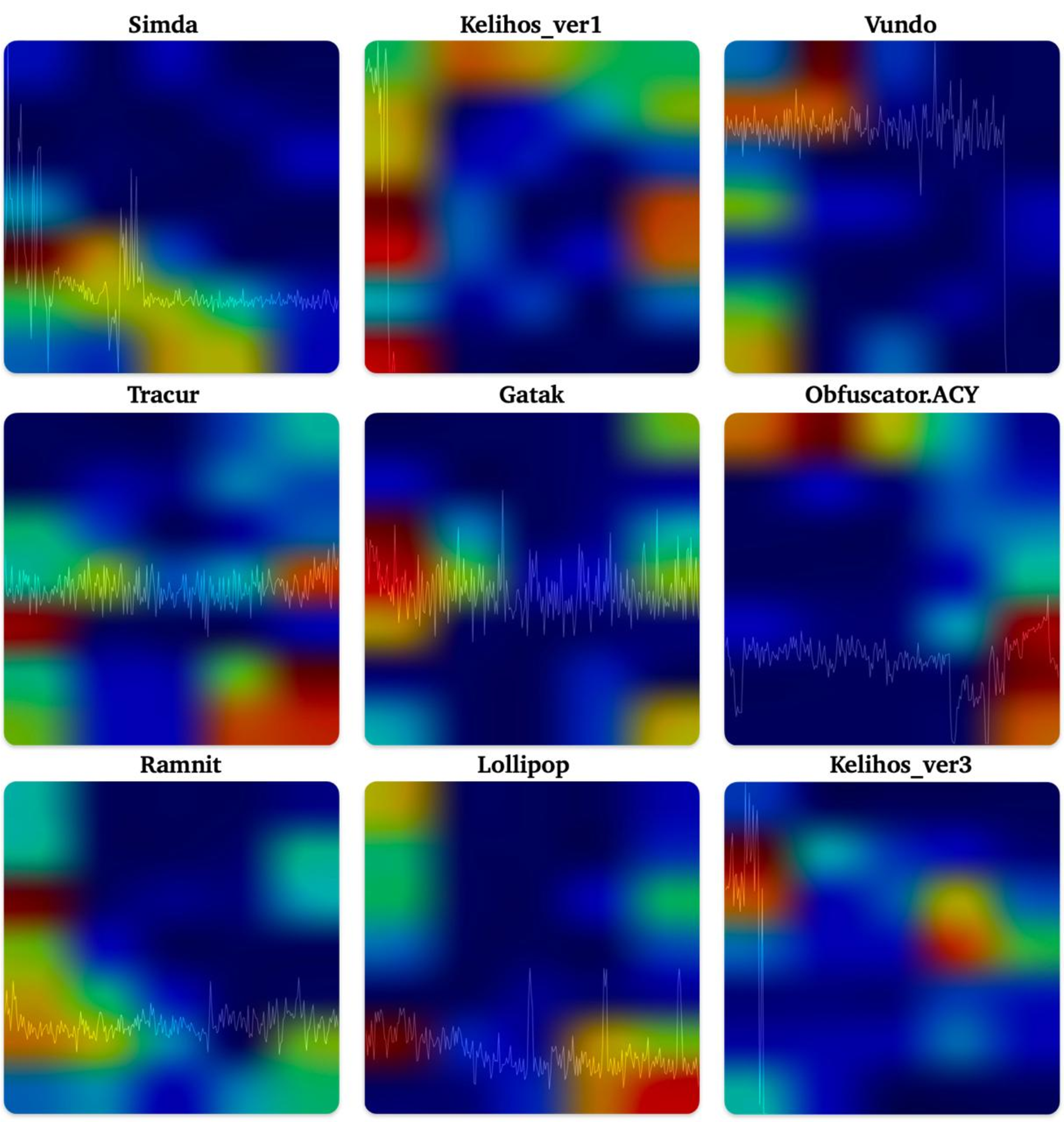}
        \caption{$EG$}
        \label{fig:gradCAM[EG]}
    \end{subfigure}
    ~ 
    \begin{subfigure}[b]{0.31\textwidth}
        \includegraphics[width=1\textwidth]{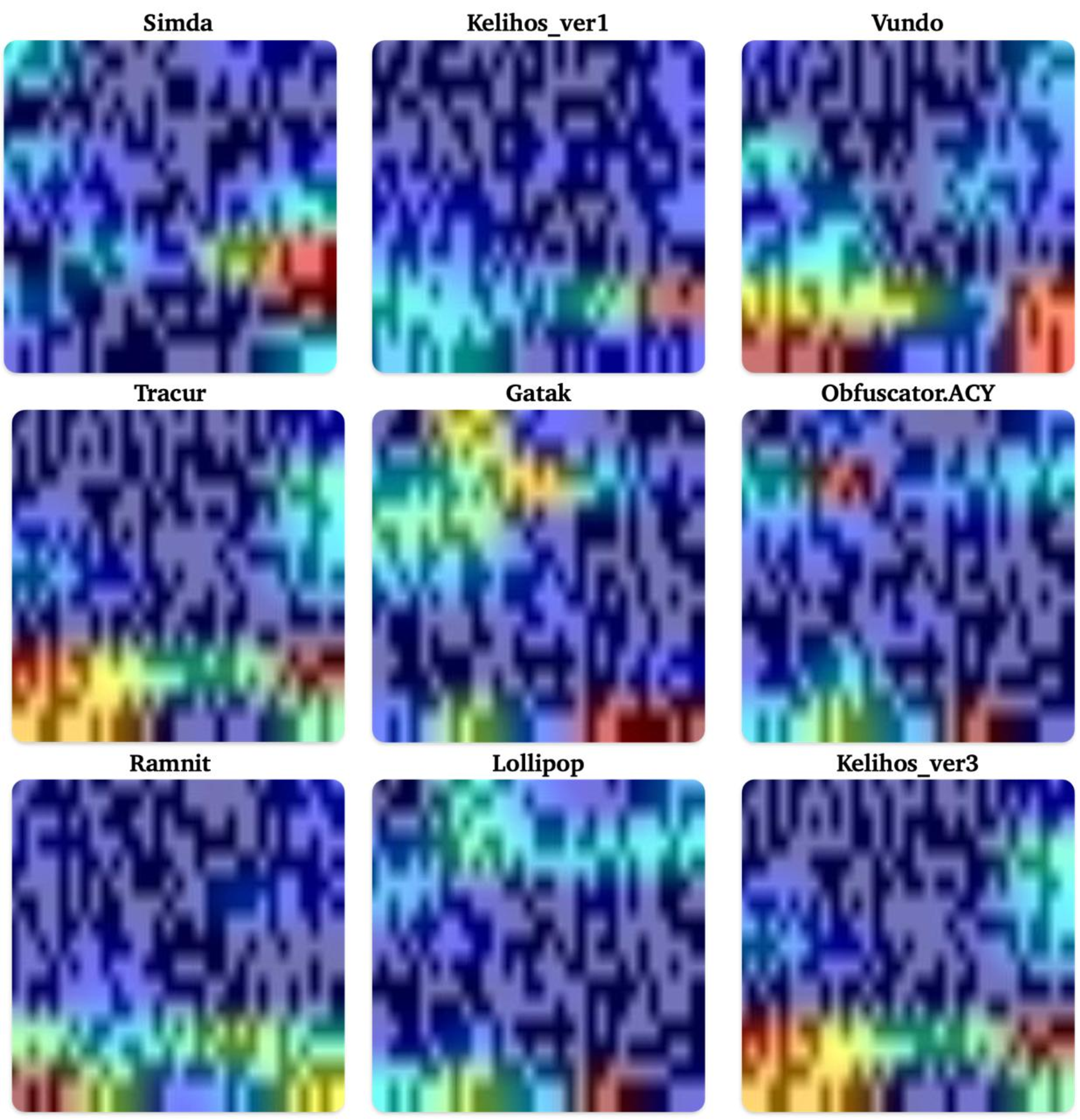}
        \caption{$SH$}
        \label{fig:gradCAM[SH]}
    \end{subfigure}
    \caption{Grad-CAM Visualization of 9 different families from the Independent Feature VGG16 Models, superimposed on, a) Grayscale image b) Entropy Graph c) Simhash Image.}\label{fig:gradCAM}
\end{figure}

\begin{figure}[!htbp]
    \centering
    \includegraphics[scale=0.093]{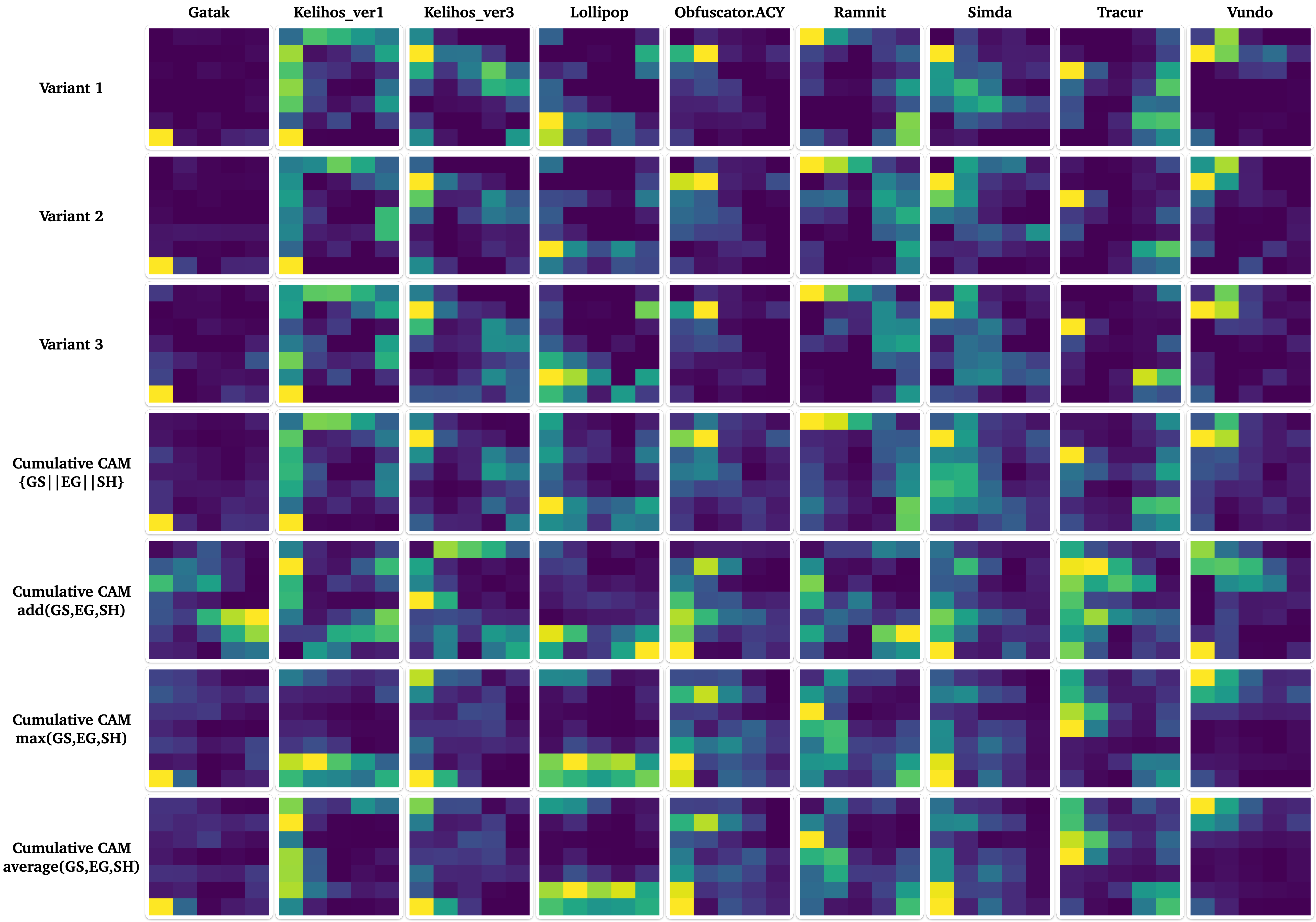}
    \caption{Grad-CAM Representation of 3 variants of 9 Malware families generated from the activation map of $GS||EG||SH$ and Cumulative Grad-CAM of $GS||EG||SH$, $add(GS, EG, SH)$, $max(GS, EG, SH)$, and $average(GS, EG, SH)$ models.}
    \label{fig:cumCAM}
\end{figure}

 The colors in an aggregated class activation map (CAM) represent the importance of different regions in the original image for a specific class prediction. The intensity of colors indicates the saliency corresponding to the pixel. The intensity of color ranges from cool to warm (blue to red) which represents the strength of different regions in activation maps. Warmer colors usually indicate higher activation or importance, while cooler colors suggest lower activation. We generate cumulative Grad-CAM images for individual VGG16 models~(\textit{Experiment-1}) trained on grayscale, entropy graph, and simhash images, as illustrated in Figure~\ref{fig:gradCAM[GS]}, Figure~\ref{fig:gradCAM[EG]}, and Figure~\ref{fig:gradCAM[SH]}. These images unveil distinctive patterns for each malware family through variations in pixel intensity, prominently observable in warmer pixels. This observation emphasizes that malware families manifest diverse structural properties and byte variability, precisely learned by CNN to enhance categorization. Additionally, in Experiment 2, we empirically validate the significance of a multimodal concatenation model for malware classification. We visualize activation maps for all malware families across three randomly selected files and cumulative class activation maps (refer Figure~\ref{fig:cumCAM}) for the entire test set submitted to different models. The \texttt{concatenation} model excels in capturing meaningful feature relationships, enabling more accurate malware classification. Clear distinctions among all nine families are evident from the pixel variations, highlighting structural similarities within each family through identical trends observed vertically.

\begin{table}[!htbp]
    \centering
    \caption{This table shows the F1-Score of individual features, \textit{GS}, \textit{EG}, \textit{SH}, and their different combinations. $||$ in the table represents the concatenation of features, $+$ represents $add(x,y)$, where $x$ and $y$ are two features, $avg()$, and $max()$ represent the average and maximum layer used in fusing the features. Families included in the table are: $F1(Lollipop), F2(Kelihos\_ver3), F3(Gatak), F4(Kelihos\_ver1), \\F5(Obfuscator.ACY), F6(Simda), F7(Tracur), F8(Vundo), F9(Ramnit)$}
    \begin{tabularx}{\textwidth}{l*{9}{X}}
        \toprule
        & & & & \textbf{Families} & & & \\ \cmidrule{2-10}
         \textbf{Models} & \textbf{F1} & \textbf{F2} & \textbf{F3} & \textbf{F4} & \textbf{F5} & \textbf{F6} & \textbf{F7} & \textbf{F8} & \textbf{F9} \\ \midrule
         $GS$ &  0.990 & 0.998 & 1.0 & 1.0 & 0.955 & 0.974 & 0.857 & 0.935 & 0.969  \\ 
         $EG$ & 0.980 & 0.996 & 0.998 & 0.995 & 0.979 & 0.991 & 1.0 & 0.996 & 1.0  \\ 
         $SH$ & 0.998 & 0.996 & 0.993 & 0.968 & 0.783 & 0.982 & 0.967 & 0.966 & 0.952  \\ 
         $GS||EG$ & 1.0 & 1.0 & 1.0 & 1.0 & 1.0 & 1.0 & 1.0 & 1.0 & 1.0  \\ 
         $GS||SH$ & 1.0 & 1.0 & 1.0 & 1.0 & 1.0 & 1.0 & 1.0 & 1.0 & 1.0 \\ 
         $EG||SH$ & 1.0 & 1.0 & 1.0 & 1.0 & 0.998 & 1.0 & 1.0 & 1.0 & 1.0  \\ 
         $GS+EG$ & 0.968 & 0.961 & 0.997 & 0.994 & 0.965 & 0.973 & 0.187 & 0.981 & 0.964  \\ 
         $GS+SH$ & 0.993 & 0.953 & 0.998 & 0.983 & 0.965 & 0.977 & 0.500 & 0.863 & 0.985  \\ 
         $EG+SH$ & 0.889 & 0.800 & 0.987 & 0.962 & 0.884 & 0.944 & 0.625 & 0.922 & 0.935 \\
         $avg(GS,EG)$ & 0.990 & 0.976 & 1.0 & 0.994 & 0.967 & 0.980 & 0.500 & 0.972 & 0.985  \\ 
         $avg(GS,SH)$ & 0.997 & 0.915 & 1.0 & 0.983 & 0.989 & 0.946 & 0.187 & 0.918 & 0.906 \\ 
         $avg(EG,SH)$ & 0.898 & 0.684 & 0.988 & 0.975 & 0.938 & 0.896 & 0.060 & 0.890 & 0.964 \\ 
         $max(GS,EG)$ & 0.993 & 0.969 & 0.998 & 0.992 & 0.967 & 0.980 & 0.250 & 0.968 & 0.964 \\ 
         $max(GS,SH)$ & 0.993 & 0.946 & 0.998 & 0.984 & 0.978 & 0.964 & 0.00 & 0.986 & 0.877  \\ 
         $max(EG,SH)$ & 0.917 & 0.800 & 0.987 & 0.961 & 0.897 & 0.862 & 0.062 & 0.954 & 0.719 \\
         $GS||EG||SH$ & 1.0 & 1.0 & 1.0 & 1.0 & 1.0 & 1.0 & 1.0 & 1.0 & 1.0 \\ 
         $GS+EG+SH$ & 0.996 & 0.915 & 0.998 & 0.994 & 0.975 & 0.986 & 0.312 & 0.959 & 0.978 \\ 
         $avg(GS,EG,SH)$ & 0.993 & 0.923 & 0.998 & 0.998 & 0.989 & 0.988 & 0.255 & 0.968 & 0.964 \\ 
         $max(GS,EG,SH)$ & 0.996 & 0.946 & 1.0 & 0.996 & 0.975 & 0.980 & 0.437 & 0.936 & 0.992 \\ \bottomrule
          
    \end{tabularx}
    \label{tab:familyclassification}
\end{table}

\begin{center}
\fcolorbox{black}{lightgray}{
\begin{minipage}[center]{0.97\linewidth}
\textbf{$RQ_3$ Summary:} \em{
The utilization of t-SNE and Grad-CAM techniques to visualize features extracted from concatenation models enhances interpretability, aiding in model trust, understanding, and highlights its superiority in capturing meaningful relationships between distinct feature patterns for accurate malware classification.
} 
\end{minipage}}
\end{center}

\subsection{Comparison with Other Researchers' Methods}
Table \ref{comparison} illustrates a comparative analysis of our proposed model with state-of-the-art approaches. Here, we specifically focus on the experiment's model, the image type, and the information source utilized for image generation, addressing time analysis, and ultimately assessing the interpretability of the model. Recent studies predominantly concentrate on deep learning models due to their ability to automatically learn relevant features, resulting in accuracy ranging from 93\% to 99.9\%. Most researchers utilize grayscale images, generated from bytes, while some use opcodes for image creation. The table reveals fewer works employing a multimodal architecture with images generated from multiple information sources. Although the author in\cite{gibert2022fusing} attempted a hybrid approach, it necessitated feature engineering as well. Furthermore, it is noteworthy that no one in Table \ref{comparison} has addressed the crucial aspect of interpretability in the context of machine learning and deep learning models. Thus, the table demonstrates the superiority of our proposed model over all other existing works.
\begin{table}[!htbp]
    \centering
    \caption{Comparative analysis of the proposed model against state-of-the-art models on the BIG 2015 dataset.(GS: Grayscale, EG: Entropy Graph, and SH: Simhash)}
   \begin{tabular}{p{4cm} p{2.5cm} c c c c c}
    \toprule
      \textit{\textbf{ Model used}} & \textit{\textbf{Image}} & \textit{\textbf{Information}} & \textit{\textbf{Accuracy}} & \textit{\textbf{F1-score}} & \textit{\textbf{Interpretability}} & \textit{\textbf{Time for }} \\ 
    \textbf{} & \textbf{} & \textit{\textbf{ to generate}} & \textbf{(\%)} & \textbf{(\%)} & \textbf{} & \textit{\textbf{analysis}} \\ 
      \textbf{} & \textbf{} & \textit{\textbf{image}} &  &  & \textbf{} & \textit{\textbf{addressed}} \\ \midrule
      \cite{narayanan2016performance}Linear KNN+PCA& GS images  & Bytes & 96.6\% & - & \xmark  & \xmark  \\ 
       Features &  & &&  &   &  \\
      \cite{tekerek2022novel}CycleGAN+\newline
        DenseNet121 & RGB images & Bytes& 99.73\% & 97.76\% & \xmark & \xmark \\ 
      \cite{darem2021visualization}CNN+ & GS images & Bytes & 99.12\% & - &  \xmark & \cmark\\ 
      XGBoost & & & & & & \\ 
      \cite{pinhero2021malware}CNN+FC & GS, RGB, & Bytes & 99.20\% & 99.20\% & \xmark & \cmark \\
      & Markov images & & & &  &  \\
      \cite{alzubi2023fusion} Fusion & GS images& Bytes & 98.21\% &  98.83\%& \xmark & \cmark \\
      (LSTM+GRU) & & & & & &\\ 
      \cite{gibert2019using} 3Conv+ 1FC & GS images & Bytes &97.5\% & 94.00\% & \xmark & \cmark  \\ 
      \cite{kalash2018malware} Pre-traiined  & GS images  & Bytes & 99.97\% & - &\xmark & \xmark  \\
    VGG16+softmax & &  &  &  & &   \\
      \cite{gibert2022fusing} Ensemble  & GS images & Bytes & 99.81\% & - &\xmark & \xmark  \\ 
        CNN (Early Fusion)&  Assembly-based  &  & &  & &   \\
        & Hexadecimal- & & & & & \\ 
        & based & & & & & \\ 
      \cite{kumar2022dtmic} Fine-tuned & GS images & Bytes & 93.19\% & - &\xmark & \xmark  \\ 
      VGG16+FC & & & & & & \\ 
      \cite{ahmadi2016novel}XGBoost & GS images  & Bytes & 96.90\% & 92.82\% & \xmark & \xmark  \\ 
        classifier & (Haralick   & & & & &  \\ 
         & features)  & & & & &  \\ 
      \cite{ahmadi2016novel}XGBoost & GS images & Bytes & 97.24\% & 95.30\% & \xmark & \xmark  \\ 
        classifier & (LBP features)  & &  & & &  \\ 
      \cite{le2018deep}CNN-BiLSTM& GS images  & Bytes &98.20\% & 96.05\% & \xmark & \cmark  \\ 
      \cite{wang2021novel}DenseNet & GS images & Bytes & 97.30\% & 95.40\% &\xmark & \xmark  \\ 
      \cite{khan2019analysis} ResNet152& GS images & Opcodes &88.36\% & - &\xmark & \cmark  \\ 
      \cite{ni2018malware}CNN  & SH images & Opcodes &98.86\% & - &\xmark & \cmark  \\ 
      \cite{sun2018deep}RNN  & GS images  & Opcodes & 99.5\% & - &\xmark & \cmark  \\ 
      \cite{qiao2019multi}LeNet5 & GS images & Bytes & 98.76\% & -  &\xmark & \xmark  \\ 
      \cite{yang2018convolutional}CNN& GS images & Bytes & 98.7\% &  - &\xmark & \xmark  \\ 
      \cite{venkatraman2019hybrid}CNN-BiGRU& GS images & Bytes & -&  71.1\% &\xmark & \xmark  \\ 
      \cite{roseline2020intelligent}Deep Forest& GS images & Bytes & 97.2\%&  97.2\% &\xmark & \cmark  \\ 
      \cite{kang2019long}LSTM & GS images & asm & 97.59\% &  - & \xmark & \xmark  \\ 
      \cite{kadri2019transfer} CNN & GS images & Bytes & 98.00\%&  - &\xmark & \xmark  \\
        Proposed  & GS, EG, SH images & Bytes, & 100\% &  100\% &\cmark & \cmark  \\ 
        concatenated  &  &  Entropy,  &  &  & t-SNE, &   \\ 
        CNN(Pretrained VGG16)  &  &  asm &  &  & Grad-CAM  &   \\  \bottomrule 
    \end{tabular}
    \label{comparison}
\end{table}


\section{Conclusion and Future Research}
\label{sec:conclusion}
In this paper, we investigated the effectiveness of utilizing the VGG16 model trained on different modalities for malware categorization. Through detailed experiments, we analyzed the influence of merging/fusion operators on classifier performance. Our thorough analysis revealed that the concatenation operator accurately detects all malware families using bi-modal or multimodal feature representation, demonstrating promising execution times akin to real-time malware scanning. This operator enhances the model's capacity to learn complex representations from diverse feature spaces of different modalities. In addition, we extend the model interpretability by visualizing class activation maps and t-SNE. Looking ahead, our research will explore the byte-code patterns learned by deep learning models, transitioning towards developing interpretable malware models. Additionally, we aim to create family signatures by fusing important regions from malware variant heatmaps. Finally, we seek to identify the regions within malware images contributing to classification, examining dispersed pixel regions from both the head and tail of the images.
\section*{Declarations}
\textbf{Conflict of interest:} All authors declare that they have no conflict of interest.
\newline
\section*{Acknowledgment}
\begin{figure}[!htbp]
     \includegraphics[scale=0.25]{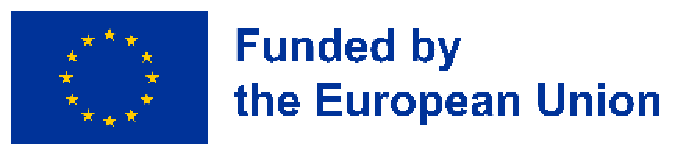}
\end{figure}
This work was supported in part by the project HORIZON Europe Framework Programme through the project ``OPTIMA - Organization sPecific Threat Intelligence Mining and Sharing'' (code: 101063107).
\bibliographystyle{plain}
\bibliography{references}

\begin{thebibliography}{10}

\bibitem{ahmadi2016novel}
Mansour Ahmadi, Dmitry Ulyanov, Stanislav Semenov, Mikhail Trofimov, and Giorgio Giacinto.
\newblock Novel feature extraction, selection and fusion for effective malware family classification.
\newblock In {\em Proceedings of the sixth ACM conference on data and application security and privacy}, pages 183--194, 2016.

\bibitem{AHMED202311}
Mumtaz Ahmed, Neda Afreen, Muneeb Ahmed, Mustafa Sameer, and Jameel Ahamed.
\newblock An inception v3 approach for malware classification using machine learning and transfer learning.
\newblock {\em International Journal of Intelligent Networks}, 4:11--18, 2023.

\bibitem{alzubi2023fusion}
Omar~A Alzubi, Issa Qiqieh, and Jafar~A Alzubi.
\newblock Fusion of deep learning based cyberattack detection and classification model for intelligent systems.
\newblock {\em Cluster Computing}, 26(2):1363--1374, 2023.

\bibitem{darem2021visualization}
Abdulbasit Darem, Jemal Abawajy, Aaisha Makkar, Asma Alhashmi, and Sultan Alanazi.
\newblock Visualization and deep-learning-based malware variant detection using opcode-level features.
\newblock {\em Future Generation Computer Systems}, 125:314--323, 2021.

\bibitem{gibert2020hydra}
Daniel Gibert, Carles Mateu, and Jordi Planes.
\newblock Hydra: A multimodal deep learning framework for malware classification.
\newblock {\em Computers \& Security}, 95:101873, 2020.

\bibitem{gibert2019using}
Daniel Gibert, Carles Mateu, Jordi Planes, and Ramon Vicens.
\newblock Using convolutional neural networks for classification of malware represented as images.
\newblock {\em Journal of Computer Virology and Hacking Techniques}, 15:15--28, 2019.

\bibitem{gibert2022fusing}
Daniel Gibert, Jordi Planes, Carles Mateu, and Quan Le.
\newblock Fusing feature engineering and deep learning: A case study for malware classification.
\newblock {\em Expert Systems with Applications}, 207:117957, 2022.

\bibitem{humeau2019texture}
Anne Humeau-Heurtier.
\newblock Texture feature extraction methods: A survey.
\newblock {\em IEEE access}, 7:8975--9000, 2019.

\bibitem{jian2021novel}
Yifei Jian, Hongbo Kuang, Chenglong Ren, Zicheng Ma, and Haizhou Wang.
\newblock A novel framework for image-based malware detection with a deep neural network.
\newblock {\em Computers \& Security}, 109:102400, 2021.

\bibitem{kadri2019transfer}
Mohamad~Al Kadri, Mohamed Nassar, and Haidar Safa.
\newblock Transfer learning for malware multi-classification.
\newblock In {\em Proceedings of the 23rd International Database Applications \& Engineering Symposium}, pages 1--7, 2019.

\bibitem{kalash2018malware}
Mahmoud Kalash, Mrigank Rochan, Noman Mohammed, Neil~DB Bruce, Yang Wang, and Farkhund Iqbal.
\newblock Malware classification with deep convolutional neural networks.
\newblock In {\em 2018 9th IFIP international conference on new technologies, mobility and security (NTMS)}, pages 1--5. IEEE, 2018.

\bibitem{kang2019long}
Jungho Kang, Sejun Jang, Shuyu Li, Young-Sik Jeong, and Yunsick Sung.
\newblock Long short-term memory-based malware classification method for information security.
\newblock {\em Computers \& Electrical Engineering}, 77:366--375, 2019.

\bibitem{khan2019analysis}
Riaz~Ullah Khan, Xiaosong Zhang, and Rajesh Kumar.
\newblock Analysis of resnet and googlenet models for malware detection.
\newblock {\em Journal of Computer Virology and Hacking Techniques}, 15:29--37, 2019.

\bibitem{kumar2022dtmic}
Sanjeev Kumar and B~Janet.
\newblock Dtmic: Deep transfer learning for malware image classification.
\newblock {\em Journal of Information Security and Applications}, 64:103063, 2022.

\bibitem{le2018deep}
Quan Le, Ois{\'\i}n Boydell, Brian Mac~Namee, and Mark Scanlon.
\newblock Deep learning at the shallow end: Malware classification for non-domain experts.
\newblock {\em Digital Investigation}, 26:S118--S126, 2018.

\bibitem{mallik2022conrec}
Abhishek Mallik, Anavi Khetarpal, and Sanjay Kumar.
\newblock Conrec: malware classification using convolutional recurrence.
\newblock {\em Journal of Computer Virology and Hacking Techniques}, 18(4):297--313, 2022.

\bibitem{martin2019android}
Alejandro Mart{\'\i}n, Ra{\'u}l Lara-Cabrera, and David Camacho.
\newblock Android malware detection through hybrid features fusion and ensemble classifiers: The andropytool framework and the omnidroid dataset.
\newblock {\em Information Fusion}, 52:128--142, 2019.

\bibitem{naeem2019identification}
Hamad Naeem, Bing Guo, Muhammad~Rashid Naeem, Farhan Ullah, Hamza Aldabbas, and Muhammad~Sufyan Javed.
\newblock Identification of malicious code variants based on image visualization.
\newblock {\em Computers \& Electrical Engineering}, 76:225--237, 2019.

\bibitem{narayanan2016performance}
Barath~Narayanan Narayanan, Ouboti Djaneye-Boundjou, and Temesguen~M Kebede.
\newblock Performance analysis of machine learning and pattern recognition algorithms for malware classification.
\newblock In {\em 2016 IEEE national aerospace and electronics conference (NAECON) and ohio innovation summit (OIS)}, pages 338--342. IEEE, 2016.

\bibitem{nataraj2011malware}
Lakshmanan Nataraj, Sreejith Karthikeyan, Gregoire Jacob, and Bangalore~S Manjunath.
\newblock Malware images: visualization and automatic classification.
\newblock In {\em Proceedings of the 8th international symposium on visualization for cyber security}, pages 1--7, 2011.

\bibitem{ni2018malware}
Sang Ni, Quan Qian, and Rui Zhang.
\newblock Malware identification using visualization images and deep learning.
\newblock {\em Computers \& Security}, 77:871--885, 2018.

\bibitem{NI2018871}
Sang Ni, Quan Qian, and Rui Zhang.
\newblock Malware identification using visualization images and deep learning.
\newblock {\em Computers \& Security}, 77:871--885, 2018.

\bibitem{pinhero2021malware}
Anson Pinhero, ML~Anupama, P~Vinod, Corrado~Aaron Visaggio, N~Aneesh, S~Abhijith, and S~AnanthaKrishnan.
\newblock Malware detection employed by visualization and deep neural network.
\newblock {\em Computers \& Security}, 105:102247, 2021.

\bibitem{qiao2019multi}
Yanchen Qiao, Qingshan Jiang, Zhenchao Jiang, and Liang Gu.
\newblock A multi-channel visualization method for malware classification based on deep learning.
\newblock In {\em 2019 18th IEEE International Conference On Trust, Security And Privacy In Computing And Communications/13th IEEE International Conference On Big Data Science And Engineering (TrustCom/BigDataSE)}, pages 757--762. IEEE, 2019.

\bibitem{ren2020malware}
Zhuojun Ren, Guang Chen, and Wenke Lu.
\newblock Malware visualization methods based on deep convolution neural networks.
\newblock {\em Multimedia Tools and Applications}, 79:10975--10993, 2020.

\bibitem{rezende2018malicious}
Edmar Rezende, Guilherme Ruppert, Tiago Carvalho, Antonio Theophilo, Fabio Ramos, and Paulo~de Geus.
\newblock Malicious software classification using vgg16 deep neural network’s bottleneck features.
\newblock In {\em Information Technology-New Generations: 15th International Conference on Information Technology}, pages 51--59. Springer, 2018.

\bibitem{roseline2020intelligent}
S~Abijah Roseline, S~Geetha, Seifedine Kadry, and Yunyoung Nam.
\newblock Intelligent vision-based malware detection and classification using deep random forest paradigm.
\newblock {\em IEEE Access}, 8:206303--206324, 2020.

\bibitem{selvaraju2017grad}
Ramprasaath~R Selvaraju, Michael Cogswell, Abhishek Das, Ramakrishna Vedantam, Devi Parikh, and Dhruv Batra.
\newblock Grad-cam: Visual explanations from deep networks via gradient-based localization.
\newblock In {\em Proceedings of the IEEE international conference on computer vision}, pages 618--626, 2017.

\bibitem{sun2018deep}
Guosong Sun and Quan Qian.
\newblock Deep learning and visualization for identifying malware families.
\newblock {\em IEEE Transactions on Dependable and Secure Computing}, 18(1):283--295, 2018.

\bibitem{tekerek2022novel}
Adem Tekerek and Muhammed~Mutlu Yapici.
\newblock A novel malware classification and augmentation model based on convolutional neural network.
\newblock {\em Computers \& Security}, 112:102515, 2022.

\bibitem{theckedath2020detecting}
Dhananjay Theckedath and RR~Sedamkar.
\newblock Detecting affect states using vgg16, resnet50 and se-resnet50 networks.
\newblock {\em SN Computer Science}, 1:1--7, 2020.

\bibitem{vasan2020imcfn}
Danish Vasan, Mamoun Alazab, Sobia Wassan, Hamad Naeem, Babak Safaei, and Qin Zheng.
\newblock Imcfn: Image-based malware classification using fine-tuned convolutional neural network architecture.
\newblock {\em Computer Networks}, 171:107138, 2020.

\bibitem{vasan2020image}
Danish Vasan, Mamoun Alazab, Sobia Wassan, Babak Safaei, and Qin Zheng.
\newblock Image-based malware classification using ensemble of cnn architectures (imcec).
\newblock {\em Computers \& Security}, 92:101748, 2020.

\bibitem{venkatraman2019hybrid}
Sitalakshmi Venkatraman, Mamoun Alazab, and R~Vinayakumar.
\newblock A hybrid deep learning image-based analysis for effective malware detection.
\newblock {\em Journal of Information Security and Applications}, 47:377--389, 2019.

\bibitem{wang2021novel}
Changguang Wang, Ziqiu Zhao, Fangwei Wang, and Qingru Li.
\newblock A novel malware detection and family classification scheme for iot based on deam and densenet.
\newblock {\em Security and Communication Networks}, 2021:1--16, 2021.

\bibitem{xiao2021image}
Mao Xiao, Chun Guo, Guowei Shen, Yunhe Cui, and Chaohui Jiang.
\newblock Image-based malware classification using section distribution information.
\newblock {\em Computers \& Security}, 110:102420, 2021.

\bibitem{yang2018convolutional}
Chun Yang, Yu~Wen, Jianbin Guo, Haitao Song, Linfeng Li, Haoyang Che, and Dan Meng.
\newblock A convolutional neural network based classifier for uncompressed malware samples.
\newblock In {\em Proceedings of the 1st Workshop on Security-Oriented Designs of Computer Architectures and Processors}, pages 15--17.

\bibitem{yuan2020byte}
Baoguo Yuan, Junfeng Wang, Dong Liu, Wen Guo, Peng Wu, and Xuhua Bao.
\newblock Byte-level malware classification based on markov images and deep learning.
\newblock {\em Computers \& Security}, 92:101740, 2020.

\bibitem{zhu2021malware}
Xuejin Zhu, Jie Huang, Bin Wang, and Chunyang Qi.
\newblock Malware homology determination using visualized images and feature fusion.
\newblock {\em PeerJ Computer Science}, 7:e494, 2021.

\end{thebibliography}
\balance

\end{document}